\documentclass{aastex701}

\usepackage{siunitx}
\DeclareSIUnit{\decibelmilliwatt}{dBm}
\DeclareSIUnit{\decibelcarrier}{dBc}
\DeclareSIUnit{\decibelisotropic}{dBi}
\DeclareSIUnit{\samplepersecond}{SPS}
\usepackage{amsmath}
\usepackage[noabbrev]{cleveref}
\usepackage{acronym}

\acrodef{DRAO}{Dominion Radio Astrophysical Observatory}
\acrodef{ANSI}{American National Standards Institute}
\acrodef{VITA}{VMEbus International Trade Association}
\acrodef{UVic}{University of Victoria}
\acrodef{NRC}{National Research Council}
\newacroindefinite{NRC}{an}{a}
\acrodef{ITU}{International Telecommunication Union}
\newacroindefinite{ITU}{an}{an}
\acrodef{ITU-R}{International Telecommunication Union Radiocommunication Sector}
\newacroindefinite{ITU-R}{an}{an}
\acrodef{ISED}{Innovation, Science and Economic Development Canada}
\acrodef{CHIME}{Canadian Hydrogen Intensity Mapping Experiment}
\acrodef{JAG}{John A. Galt telescope}
\acrodef{DVA}{Dish Verification Antenna}
\acrodef{SKA}{Square Kilometer Array}
\newacroindefinite{SKA}{an}{a}
\acrodef{LWA}{Long Wavelength Array}
\newacroindefinite{LWA}{an}{a}
\acrodef{SFM}{Solar Flux Monitor}
\newacroindefinite{SFM}{an}{a}
\acrodef{DDCP}{Digital Down Converter Pool}
\acrodef{USRP}{Universal Software Radio Peripheral}
\acrodef{ST}{Synthesis Telescope}
\newacroindefinite{ST}{an}{a}
\acrodef{CGEM}{Canadian Galactic Emission Mapper}
\acrodef{ARTTA}{Advanced Radio Telescope Timing Array}
\acrodef{CHORD}{Canadian Hydrogen Observatory and Radio-transient Detector}

\acrodef{UDP}{User Datagram Protocol}
\acrodef{TCP}{Transmission Control Protocol}
\acrodef{IP}{Internet Protocol}
\newacroindefinite{IP}{an}{an}
\acrodef{SFP}{small form-factor pluggable}
\newacroindefinite{SFP}{an}{a}
\acrodef{QSFP}{quad small form-factor pluggable}
\acrodef{FPGA}{field-programmable gate array}
\newacroindefinite{FPGA}{an}{a}
\acrodef{CPU}{central processing unit}
\acrodef{GPU}{graphics processing unit}
\acrodef{COTS}{commercial off-the-shelf}
\acrodef{HPC}{high-performance computing}
\newacroindefinite{HPC}{an}{a}
\acrodef{DPDK}{Data Plane Development Kit}
\acrodef{ZeroMQ}{Zero Message Queue}
\acrodef{SigMF}{Signal Metadata Format}

\acrodef{IP1dB}{input power 1~dB compression point}
\newacroindefinite{IP1dB}{an}{an}
\acrodef{OP1dB}{output power 1~dB compression point}
\newacroindefinite{OP1dB}{an}{an}
\acrodef{PSD}{power spectral density}
\acrodef{ACM}{Array Covariance Matrix}
\newacroindefinite{ACM}{an}{an}
\acrodef{EMC}{Electromagnetic Compatibility}
\newacroindefinite{EMC}{an}{an}
\acrodef{EMI}{Electromagnetic Interference}
\newacroindefinite{EMI}{an}{an}
\acrodef{RF}{radio frequency}
\newacroindefinite{RF}{an}{a}
\acrodef{SDR}{software-defined radio}
\newacroindefinite{SDR}{an}{a}
\acrodef{RFI}{radio frequency interference}
\newacroindefinite{RFI}{an}{a}
\acrodef{BW}{bandwidth}
\acrodef{RBW}{resolution bandwidth}
\newacroindefinite{RBW}{an}{a}
\acrodef{SNR}{Signal-to-Noise Ratio}
\newacroindefinite{SNR}{an}{a}
\acrodef{LFM}{Linear Frequency Modulation}
\newacroindefinite{LFM}{an}{a}
\acrodef{CW}{Continuous Waveform}
\acrodef{HF}{High Frequency}
\newacroindefinite{HF}{an}{a}
\acrodef{FFT}{Fast Fourier Transform}
\newacroindefinite{FFT}{an}{a}
\acrodef{ADC}{analog-to-digital converter}
\newacroindefinite{ADC}{an}{an}
\acrodef{STFT}{Short-time Fourier Transform}
\newacroindefinite{STFT}{an}{a}
\acrodef{IF}{intermediate frequency}
\newacroindefinite{IF}{an}{an}
\acrodef{DSP}{digital signal processing}
\acrodef{DDC}{Digital Down Converter}
\acrodef{ARX}{Analog Receiver}
\newacroindefinite{ARX}{an}{an}
\acrodef{LNA}{low-noise amplifier}
\newacroindefinite{LNA}{an}{a}
\acrodef{TTL}{transistor-transistor logic}
\acrodef{VNA}{vector network analyzer}
\acrodef{NFA}{noise-figure analyzer}
\newacroindefinite{NFA}{an}{a}
\acrodef{SFDR}{spurious-free dynamic range}
\newacroindefinite{SFDR}{an}{a}

\acrodef{TEA}{thermoelectric assembly}
\acrodef{GPIO}{general purpose input/output}
\acrodef{OSPFB}{oversampled polyphase filterbank}
\newacroindefinite{OSPFB}{an}{an}
\acrodef{FIFO}{first in, first out}
\acrodef{FSPL}{free space path loss}
\newacroindefinite{FSPL}{an}{a}
\usepackage{subcaption}
\usepackage{graphicx}
\usepackage{hyperref}

\begin{document}

\title{Wideband RFI Monitor Requirements, Design, and Commissioning at DRAO}

\correspondingauthor{Nicholas Bruce}

\author[0000-0001-8923-3828]{Nicholas Bruce}
\affiliation{Dominion Radio Astrophysical Observatory, Herzberg Astronomy and Astrophysics Research Centre, National Research Council Canada, Penticton, BC, Canada}
\affiliation{Department of Electrical and Computer Engineering, University of Victoria, Victoria, BC, Canada}
\email[show]{nicholas.bruce@nrc-cnrc.gc.ca}

\author{Charl Baard}
\affiliation{Dominion Radio Astrophysical Observatory, Herzberg Astronomy and Astrophysics Research Centre, National Research Council Canada, Penticton, BC, Canada}
\email{charl.baard@nrc-cnrc.gc.ca}

\author{Stephen Harrison}
\affiliation{Dominion Radio Astrophysical Observatory, Herzberg Astronomy and Astrophysics Research Centre, National Research Council Canada, Penticton, BC, Canada}
\affiliation{Department of Electrical and Computer Engineering, University of Victoria, Victoria, BC, Canada}
\email{msteveharrison@gmail.com}

\author{Mohammad Islam}
\affiliation{Dominion Radio Astrophysical Observatory, Herzberg Astronomy and Astrophysics Research Centre, National Research Council Canada, Penticton, BC, Canada}
\affiliation{Department of Computer Science, Math, Physics, \& Statistics, University of British Columbia, Okanagan Campus, Kelowna, BC, Canada}
\email{mohammad.islam@nrc-cnrc.gc.ca}

\author[0000-0001-9827-4533]{Abraham J. Otto}
\affiliation{Dominion Radio Astrophysical Observatory, Herzberg Astronomy and Astrophysics Research Centre, National Research Council Canada, Penticton, BC, Canada}
\email{braam.otto@nrc-cnrc.gc.ca}

\author[0009-0001-8007-1890]{Dustin Lagoy}
\affiliation{Dominion Radio Astrophysical Observatory, Herzberg Astronomy and Astrophysics Research Centre, National Research Council Canada, Penticton, BC, Canada}
\email{dustin.lagoy@nrc-cnrc.gc.ca}

\author[0009-0006-8455-6568]{Robert Messing}
\affiliation{Dominion Radio Astrophysical Observatory, Herzberg Astronomy and Astrophysics Research Centre, National Research Council Canada, Penticton, BC, Canada}
\email{rob.messing@nrc-cnrc.gc.ca}

\author{Benoit Robert}
\affiliation{Dominion Radio Astrophysical Observatory, Herzberg Astronomy and Astrophysics Research Centre, National Research Council Canada, Penticton, BC, Canada}
\email{benoit.robert@nrc-cnrc.gc.ca}

\author[0000-0002-4217-5138]{Timothy Robishaw}
\affiliation{Dominion Radio Astrophysical Observatory, Herzberg Astronomy and Astrophysics Research Centre, National Research Council Canada, Penticton, BC, Canada}
\affiliation{Department of Computer Science, Math, Physics, \& Statistics, University of British Columbia, Okanagan Campus, Kelowna, BC, Canada}
\email{tim.robishaw@nrc-cnrc.gc.ca}

\author[0000-0003-1112-3738]{Peter F. Driessen}
\affiliation{Department of Electrical and Computer Engineering, University of Victoria, Victoria, BC, Canada}
\email{peter@ece.uvic.ca}

\begin{abstract}
    In this paper, we introduce the radio frequency interference monitor deployed at the Dominion Radio Astrophysical Observatory. It provides \qty{2}{\giga\hertz} of instantaneous bandwidth, supporting channel bandwidths as fine as $\approx$\qty{100}{\hertz} for $\qty{1}{\second}$ integrations, or integration times as low as $\approx\qty{50}{\milli\second}$ for the standard \qty{3.33}{\kilo\hertz} channel bandwidth. After operating as a prototype instrument for several years, the monitor was commissioned to improve the calibration method, analog section temperature, and gain stability. It now operates both as a transient detector and as a long-term radio environment characterization tool. We introduce novel applications for the monitor and derive a new method for calculating the effect of gain drift on integrated data. 
\end{abstract}

\section{Introduction} \label{sec:introduction}

To operate Canada's \ac{DRAO} effectively, we must monitor the primary frequency bands of all operating telescopes on the site for local sources of \ac{RFI}. This monitoring effort primarily aims to preserve the radio-quiet environment of the observatory site for current and future science instruments. The \ac{DRAO} hosts leading-edge instruments owned by the Canadian government and external collaborators (see \S~\ref{sec:frequency-coverage}), all of whom directly benefit from this \ac{RFI} monitoring effort.

\begin{figure}[htb!]
    \centering
    \includegraphics{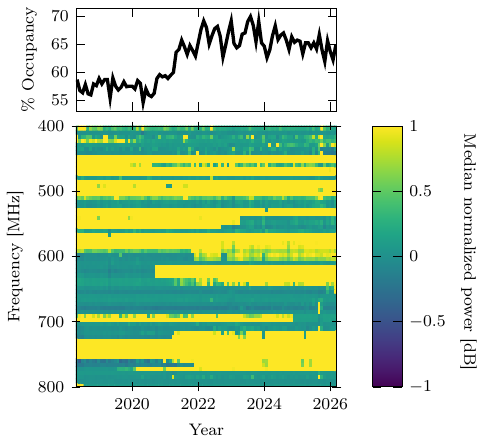}
    \caption{Annual snapshots of \ac{CHIME} data normalized against a snapshot median (lower) showing an increasing trend in percent occupancy above \qty{0}{\decibel} (upper) \citep{Amiri_2022}.}
    \label{fig:chime}
\end{figure}

The monitor serves the science community by confirming interference events in observation data. It further enables study of the statistical nature of the local \ac{RF} environment to establish the suitability of proposed experiments. To complete these feasibility assessments, accurate long-term statistics characterizing the \ac{DRAO} \ac{RF} environment are crucial. Instruments from external collaborators, such as the \ac{CHIME}, can provide insight into \ac{RF} environment trends on site (\Cref{fig:chime}), but they monitor only their specific frequency ranges of interest, are not typically designed to identify \ac{RFI} sources, lack shared noise-floor characteristics with other instruments, and have finite operational lifespans. Consequently, a dedicated instrument is valuable for identifying long-term trends on site. A standardized comparison of spectra across multiple years is possible only if the instrument is well calibrated to remove the effects of instabilities, yielding accurate power measurements.

The monitor assists the local operations team in their daily effort to preserve the observatory's radio-quiet environment. Meanwhile, instrument builders rely on these calibrated power measurements to design hardware suited for typical on-site power levels. Spectrum managers leverage the data to report accurate spectrum utilization and interference statistics to national and international regulators, including \ac{ISED} and the \ac{ITU}. Finally, the monitor system serves as a dedicated platform for researching advanced \ac{RFI} detection and mitigation algorithms.

The monitor initially operated as a prototype for five years \citep{harrison_rfi_2021}, during which time it streamed real-time data that detected various interfering sources, such as handheld cellular devices carried by delivery drivers, telematic data signals from vehicles parked on site, and an unshielded oscillator on a local telescope antenna. It also served as a foundational platform for related research, such as applying synthesis filterbanks to the problem of \ac{RFI} identification \citep{bruce_nicholas_ddc_2022} and performing unsupervised signal modulation recognition \citep{bruce_nicholas_modrec_2024}.

\begin{figure}[htb!]
    \centering
    \includegraphics{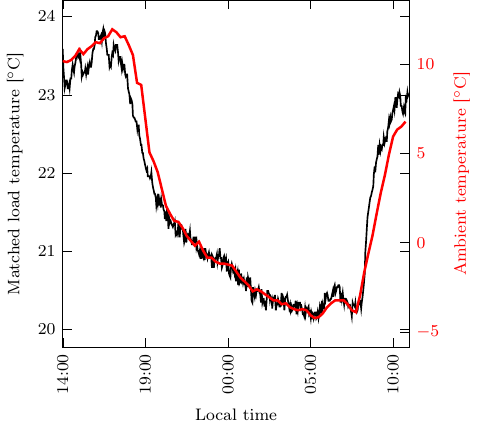}
    \caption{Analog signal chain temperature (black) and ambient temperature (red) during a day in mid-March, when ambient fluctuations were small relative to those in the summer months.}
    \label{fig:matched-load-temp-tracking-ambient}
\end{figure}

While the monitor operated as a prototype, telemetry data revealed that the analog signal chain---which must remain stable for accurate calibration---fluctuated with the outdoor temperature (\Cref{fig:matched-load-temp-tracking-ambient}) and that the heater-cooler unit inside the \ac{RF} enclosure was operating at maximum capacity.

\begin{figure}[htb!]
    \centering
    \includegraphics{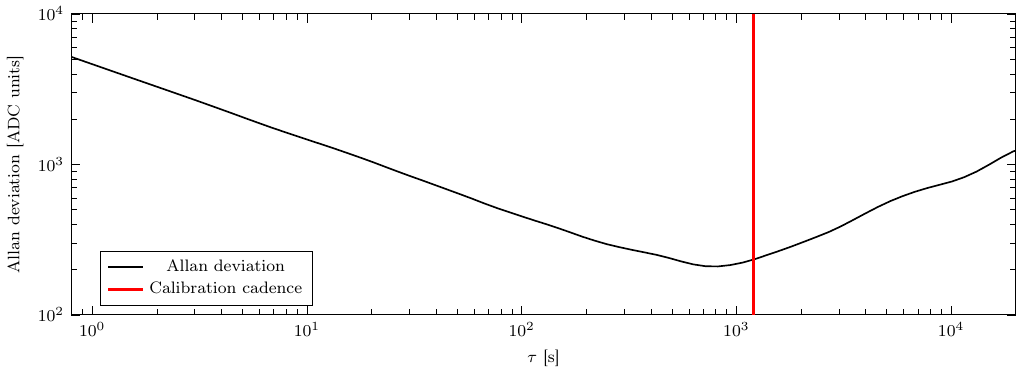}
    \caption{Allan deviation (black) at \qty{1250}{\mega\hertz} demonstrating a stability limit at $\approx\qty{750}{\second}$, beyond which systematic gain fluctuations dominate. The prototype calibration cadence of \qty{1200}{\second} (red) exceeded this useful integration window.}
    \label{fig:before-changes-allan-deviation}
\end{figure}

To assess the gain stability of the prototype, we computed the overlapping Allan deviation \citep{riley_handbook}. This metric requires a contiguous, uninterrupted time series, so we disabled the calibration cycle, replaced the antenna with a matched load, and collected \qty{21}{\hour} of continuous data. \Cref{fig:before-changes-allan-deviation} shows the overlapping Allan deviation and the minimum calibration cadence of \qty{1200}{\second} required to meet the on-sky specification. The Allan deviation characterizes the fractional gain stability of the system as a function of integration time, $\tau$. On the downward-sloping portion of the curve, the Allan deviation decreases as $\tau^{-1/2}$, consistent with white noise averaging down with integration time. Beyond the knee near $\tau \approx \qty{750}{\second}$, the deviation increases with $\tau$, indicating that systematic gain fluctuations dominate on timescales longer than $\tau_{\rm knee}$. Because the calibration cadence was longer than $\tau_{\rm knee}$, we could not continuously integrate to reduce the noise variance in the monitor data. Consequently, we redesigned the monitor to meet the gain stability requirements.

In this paper, we present the design, construction, and validation of \iac{RFI} monitor, emphasizing system gain stability for long-term \ac{RF} environment characterization. In \S~\ref{sec:requirements}, we present the requirements for the \ac{RFI} monitor and detail their derivation; in \S~\ref{sec:system-design}, we introduce the final design of the \ac{DRAO} \ac{RFI} monitor; and in \S~\ref{sec:validation}, we  validate the system against these requirements. Finally, Appendix \ref{app:redesign} details the decisions and justifications that drove the monitor's evolution from a prototype to a commissioned instrument.

The contributions in this paper are:
\begin{enumerate}
    \item a rigorous requirements derivation for a wideband \ac{RFI} monitoring system,
    \item a novel formulation for the effect of time-varying gain drift on integrated power measurements,
    \item a description of the commissioned \ac{DRAO} \ac{RFI} monitor design, and
    \item validation measurements demonstrating that the system meets its requirements.
\end{enumerate}

\section{System Requirements} \label{sec:requirements}

\subsection{Frequency Coverage} \label{sec:frequency-coverage}

\begin{figure}[htb!]
    \centering
    \includegraphics{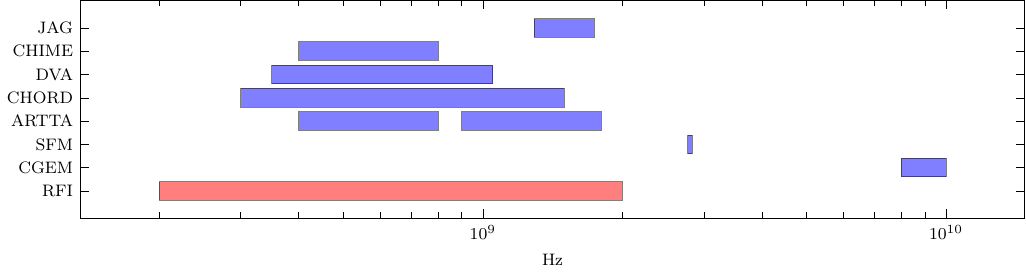}
    \caption{Frequency ranges of \ac{DRAO} telescopes along with the \ac{RFI} monitor frequency coverage. The primary frequency range of the \ac{RFI} monitor covers the majority of current and proposed instruments.}
    \label{fig:drao-freq-ranges}
\end{figure}

The primary use case for the \ac{DRAO} \ac{RFI} monitor is to provide continuous coverage from \qty{250}{\mega\hertz} up to \qty{2}{\giga\hertz}, with defined performance targets from \qty{350}{\mega\hertz} to \qty{1800}{\mega\hertz}. This frequency range covers most of the operating instruments on site, as shown in \Cref{fig:drao-freq-ranges}. These instruments include the \qty{26}{\metre} \acl{JAG} (\acs{JAG}\acused{JAG}; \qtyrange{1290}{1740}{\mega\hertz}), \ac{CHIME} (\qtyrange{400}{800}{\mega\hertz}), the \acl{DVA} (\acs{DVA}\acused{DVA}; \qtyrange{350}{1050}{\mega\hertz}), the \acl{CHORD} (\acsu{CHORD}; \qtyrange{300}{1500}{\mega\hertz}), the \acl{ARTTA} (\acs{ARTTA}\acused{ARTTA}; \qtyrange{400}{800}{\mega\hertz}, \qtyrange{900}{1800}{\mega\hertz}), the \acl{SFM} (\acs{SFM}\acused{SFM}; \qtyrange{2770}{2830}{\mega\hertz}), and the \acl{CGEM} (\acs{CGEM}\acused{CGEM}; \qtyrange{8}{10}{\giga\hertz}). Although continuous monitoring is a more costly and computationally demanding solution than frequency scanning, we assert that it is essential for characterizing both the continuous and transient local \ac{RF} environment at the desired level of detail, consistency, and sensitivity.

\subsection{System Temperature} \label{sec:sys-temp-requirement}

Unlike a highly directional radio telescope, the monitor observes a broad 
environment consisting of $\approx$60\% ground and surrounding hills and $\approx$40\% sky, a ratio derived from the site's average horizon angle of \qty{12.5}{\degree} (see Appendix \ref{app:sky-fraction}). According to \ac{ITU-R} Recommendation P.372-17, the cosmic background noise temperature is \qty{2.7}{\kelvin} and the nominal surface temperature is \qty{288}{\kelvin} \citep{itu372}. Weighting these temperatures by the monitor's spatial coverage yields an expected antenna noise temperature of $T_{\rm ant} \approx \qty{174}{\kelvin}$. Adopting a slightly conservative value of $T_{\rm ant} = \qty{170}{\kelvin}$, we specified a matching target noise figure for the analog receiver of \qty{2}{\decibel} ($T_{\rm rec}\leq$\qty{170}{\kelvin}). Consequently, the target system temperature for the instrument is $T_{\rm sys} = T_{\rm ant}+T_{\rm rec} = \qty{340}{\kelvin}$.

\subsection{Antenna Position} \label{sec:antenna-position-req}

The analog receiver chain must possess sufficient dynamic range to detect low-level interference during quiet periods while avoiding saturation from intense, highly localized \ac{RFI} generated by on-site human activity. Because staff, contractors, couriers, and visitors inevitably bring \ac{RFI}-generating devices on site,  the \ac{RFI} monitor must be positioned such that a cellular device---typically the highest-power emitter at the \ac{DRAO}---transmitting from any publicly accessible location cannot saturate the receiver.

To derive the required physical standoff distance, we establish the following parameters and constraints:
\begin{enumerate}
    \item The most powerful localized transmitters are likely to be cellular uplinks operating in the \qtyrange{824}{849}{\mega\hertz} band (centered at \qty{836.5}{\mega\hertz}) with a maximum radiated power of \qty{1}{\watt} ($P_{\rm tx}=\qty{30}{\decibelmilliwatt}$).
    \item To strictly limit nonlinearities, the \ac{RF} signal chain requires at least a \qty{10}{\decibel} safety margin below the \ac{IP1dB}.
    \item Prior to any programmable attenuation, the analog front end incorporates two cascaded, non-programmable \acp{LNA}, each providing \qty{22}{\decibel} of gain with a \qty{1.25}{\decibel} noise figure at \qty{1}{\giga\hertz}.
\end{enumerate}

The primary design constraint is preventing saturation of the second-stage \ac{LNA}. At \qty{836.5}{\mega\hertz}, this amplifier has \iac{OP1dB} of \qty{19}{\decibelmilliwatt} and \iac{IP1dB} of \qty{-2}{\decibelmilliwatt}. To maintain the \qty{10}{\decibel} safety margin, the input power to this second stage must not exceed \qty{-12}{\decibelmilliwatt}. Accounting for the \qty{22}{\decibel} gain of the first-stage \ac{LNA}, the maximum tolerable signal power arriving at the input of the receiver is $P_{\rm target} = \qty{-2}{\decibelmilliwatt}-\qty{10}{\decibel}-\qty{22}{\decibel} = \qty{-34}{\decibelmilliwatt}$.

Consequently, the physical placement of the monitor antenna must guarantee a minimum path loss of $L_{\rm target}=P_{\rm tx}-P_{\rm target}=\qty{64}{\decibel}$.

\subsection{On-Sky Time} \label{sec:on-sky-requirement}

\ac{ITU-R} Recommendation RA.1513-2 states that a radio astronomy observation is considered degraded when \ac{RFI} causes a 5\% data loss \citep{itu1513-2}. Because the monitor is specifically designed to observe \ac{RFI}, we invert this metric: data loss occurs whenever the instrument is not actively recording the \ac{RF} environment. To ensure a comprehensive long-term analysis, we therefore require the monitor to maintain a 95\% ``on-sky'' time.

\subsection{Gain Stability} \label{sec:gain-stability-requirement}

Because of highly variable ambient temperatures at the \ac{DRAO}, temperature-induced gain instability is a primary design concern. Contributors to this instability include the temperature instability of the signal chain, the noise contribution of the receiver, the duration of each calibration, and the cadence of the calibration cycle. 

Up to this point, signal levels have been characterized by their total power, $P$ (\unit{\watt}). However, evaluating wideband sensitivity requires a  transition to \ac{PSD}, $S_x$ (\unit{\watt\per\hertz}). Following the Johnson-Nyquist relationship, the total noise power in a channel of bandwidth $\Delta f$ (\unit{\hertz}) is $P=k T_{\rm sys} \Delta f$, where $k$ is Boltzmann's constant and $T_{\rm sys}$ is the system noise temperature. The radiometer input power level is therefore the \ac{PSD}, given by $S_x = P / \Delta f = k T_{\rm sys}$.

The sensitivity of a total-power radiometer employing a square-law detector\footnote{Our system records complex baseband voltages and calculates the squared magnitude to yield the signal power.} is defined as the smallest detectable change in the input \ac{PSD} level, $\Delta S_x$ (\unit{\watt\per\hertz}). This theoretical limit is governed by the ideal radiometer equation, which bounds the fractional sensitivity by the system's integration time and channel bandwidth:
\begin{equation}
    \label{eq:radiometer}
    \frac{\Delta S_x}{S_x} = \frac{\Delta T_{\rm thermal}}{T_{\rm sys}} = \frac{1}{\sqrt{\Delta f \tau}} = \frac{1}{\sqrt{N}}\,,
\end{equation}
where $\tau$ is the total integration time in seconds, $N=\Delta f \tau$ is the bandwidth-time product (equivalent to the number of independent samples collected), and $\Delta T_{\rm thermal}$ is the equivalent temperature variation in the system due to thermal noise.

To determine the maximum allowable gain change that maintains system sensitivity, we adopt the criterion from \ac{ITU-R} Recommendation RA.769-2 \citep{itu769}. This recommendation specifies a maximum 10\% error when measuring the sensitivity limit, $\Delta S_x$. Applying this tolerance yields a maximum allowable \ac{PSD} drift of
\begin{equation}
    \label{eq:delta-P-max}
    \Delta S_{x,~\rm max} = 0.1 \frac{S_x}{\sqrt{N}} = 0.1\frac{kT_{\rm sys}}{\sqrt{N}}\,.
\end{equation}

The duration over which the monitor must remain gain-stable is dictated by the calibration cadence. For the prototype, a 95\% on-sky requirement (\S~\ref{sec:on-sky-requirement}) and a \qty{60}{\second} calibration duration (Appendix~\ref{app:redesign}) limit the system to a maximum of three calibrations per hour. Consequently, the monitor must maintain gain stability over an interval of \qty{1200}{\second}.

To derive a specific requirement for the maximum noise variance attributable to gain drift (motivated by the RA.769-2 requirement for maximum interferer power), we adopt the assumptions outlined in \citet{burke_introduction_2019}:
\begin{itemize}
    \item Thermal and gain variations are independent and add quadratically:
    \begin{equation} \label{eq:thermal-effects-add-quadratically}
        \Delta T_{\rm total}^2 = \Delta T^2_{\rm thermal} + \Delta T^2_{\rm gain}\,,
    \end{equation}
    where $\Delta T_{\rm total}$ is the total noise uncertainty added to $T_{\rm sys}$ as a function of time and $\Delta T_{\rm gain}$ is the uncertainty due to gain drift.
    
    \item A fractional change in the system gain produces a proportional uncertainty in the system noise temperature,
    \begin{equation} \label{eq:T-gain-is-proportional-to-T-sys-times-gain-variation}
        \Delta T_{\rm gain} = \frac{\Delta G}{G}T_{\rm sys}\,.  
    \end{equation}
\end{itemize}

We therefore require that the noise uncertainty in the system attributable to gain drift, $\Delta T_{\rm gain}$, be no greater than 10\% of the total noise uncertainty, $\Delta T_{\rm total}$, such that
\begin{equation} \label{eq:fundamental-gain-stability-requirement}
    \Delta T_{\rm gain} \leq 0.1\Delta T_{\rm total}\,.
\end{equation}
Solving \cref{eq:thermal-effects-add-quadratically} for $\Delta T_{\rm thermal}$ and substituting into \cref{eq:fundamental-gain-stability-requirement}, we find that
\begin{equation} \label{eq:simplified-gain-uncertainty-requirement}
    \Delta T_{\rm gain} \leq 0.1\Delta T_{\rm thermal}\,.
\end{equation}
A full derivation of \cref{eq:simplified-gain-uncertainty-requirement} is provided in Appendix~\ref{app:gain-stability-requirement-algebra}. We derive a simple and measurable relationship from this requirement by substituting \cref{eq:T-gain-is-proportional-to-T-sys-times-gain-variation} into \cref{eq:simplified-gain-uncertainty-requirement}, yielding
\begin{equation*}
    \frac{\Delta G}{G} T_{\rm sys} \leq 0.1\Delta T_{\rm thermal}\,.
\end{equation*}
Dividing by $T_{\rm sys}$ gives
\begin{equation*}
    \frac{\Delta G}{G} \leq 0.1\frac{\Delta T_{\rm thermal}}{T_{\rm sys}}\,.
\end{equation*}
Applying the radiometer \cref{eq:radiometer} yields
\begin{equation*}
    \frac{\Delta G}{G} \leq 0.1 \frac{1}{\sqrt{N}}\,.
\end{equation*}
Rearranging these terms provides the final requirement,
\begin{equation} \label{eq:gain-uncertainty-requirement}
    \frac{\Delta G}{G}\sqrt{N} \leq 0.1\,.
\end{equation}

While \cref{eq:T-gain-is-proportional-to-T-sys-times-gain-variation} is standard in the literature \citep{alma992788433807291,burke_introduction_2019}, it treats gain variation as a static quantity, without accounting for how gain evolves during an integration. Because thermal variance decreases as $1/\sqrt{N}$ while gain drift accumulates over time, \cref{eq:thermal-effects-add-quadratically,eq:gain-uncertainty-requirement} do not accurately describe the stability of an integrated measurement. To address this, we define the noise uncertainty due to gain variation as the effect of the integrated gain on the output data:
\begin{equation}
    \label{eq:integrated-T-gain}
    \Delta T_{\rm gain,integrated} = \frac{T_{\rm sys}}{\tau}\int^\tau_{0}\frac{\Delta G(t)}{G(0)}dt = \frac{T_{\rm sys}}{\tau}\int^\tau_{0}\frac{G(t)-G(0)}{G(0)}dt\,,
\end{equation}
where  $G(0)$ is the initial system gain at time $t=0$, $G(t)$ is the gain at time $t$, and $\tau$ is the integration time.  By substituting \cref{eq:integrated-T-gain} into \cref{eq:simplified-gain-uncertainty-requirement} and repeating the algebraic steps used to derive \cref{eq:gain-uncertainty-requirement}, we establish a more rigorous constraint for the \ac{RFI} monitor:
\begin{equation} \label{eq:integrated-gain-uncertainty-requirement}
    \frac{\sqrt{N}}{\tau}\int^\tau_{0}\frac{G(t)-G(0)}{G(0)}dt \leq 0.1\,.
\end{equation}
\subsection{Autonomous Operation}

The monitor must autonomously collect calibrated \ac{RF} measurements and populate a continuously updated database of \ac{RFI} detections. The system requires no manual intervention for data collection or processing, completely eliminating the standard astronomical paradigms of observation planning and discrete measurement sets. Users query a single database interface, regardless of their specific scientific or engineering objectives. To enable the derivation of conventional metrics such as spectral occupancy \citep{itu2256}, the backend must compute, at a minimum, a time series of integrated power spectra.

\subsection{Software Modularity}

To provide utility beyond that of a standard radiometer, the monitor is designed as \iac{SDR}. This architecture enables advanced site characterization using techniques that operate on complex-baseband waveforms or amplitude envelopes, such as automatic modulation classification, novelty detection, and \ac{RF} transmitter fingerprinting. Consequently, the system transcends basic spectrum monitoring to function as a comprehensive platform for \ac{RF} environment analysis.

\subsection{Standard Interfaces}

Finally, standardized connectivity is critical for an open research platform. All subsystems must interconnect using established hardware and protocols. To this end, we have specified \ac{VITA} 49.0 over \ac{UDP}/\ac{IP} as the radio transport protocol for time-domain streams \citep{vita49}. Data transport is routed over \qty{10}{\giga\bit} Ethernet using \ac{QSFP} connectors, which allows the system to interface directly with commercially available hardware.

\section{System Design} \label{sec:system-design}

\subsection{Sensitivity} \label{sec:sensitivity-design}

While no formal sensitivity requirement was established, \iac{RF} link budget guided component selection to maximize the sensitivity of the monitor. Although \iac{RFI} monitor cannot practically match the sensitivity of a dedicated radio telescope, benchmarking against established astronomical standards provides a necessary target for system performance.

\ac{ITU-R} Recommendation RA.769-2 provides fundamental protection criteria for radio astronomy \citep{itu769}.  To establish the most stringent possible performance baseline, we look to the \ac{SKA}, which will be the world's largest and most sensitive radio telescope. Because it requires protection for the deepest possible observations, the \ac{SKA} \ac{RFI}/\ac{EMC} Standard \citep{skaemi4} builds upon the \citet{itu769} methodologies to derive even stricter protection levels for both continuum and spectral-line observations. Expressed as \ac{PSD} (\unit{\decibelmilliwatt\per\hertz}), these limits are
\begin{equation}
    S_x(f_c)=\begin{cases}
        -17\log_{10}(f_c)-192, & \text{continuum for~~~\qtyrange{50}{2000}{\mega\hertz}}\,, \\
        -17\log_{10}(f_c)-177, & \text{spectral line for~\qtyrange{50}{2000}{\mega\hertz}}\,,
    \end{cases}
    \label{eq:telescope-protection-levels} 
\end{equation}
where $f_c$ is the center frequency of an observation in megahertz. These protection levels assume fractional channel bandwidths of 1\% of $f_c$ for continuum observations and 0.001\% of $f_c$ for spectral-line observations.

Because the radiometer equation demonstrates that noise temperature uncertainty scales proportionally to $1/\sqrt{\Delta f}$, the sensitivity difference between these two observation modes is entirely dictated by the ratio of their bandwidths. The resulting \qty{15}{\decibel} offset between the continuum and spectral-line protection levels is calculated directly as
\begin{equation}
    10\log_{10}\left(\sqrt{\frac{\Delta f_{\rm continuum}}{\Delta f_{\rm spectral~line}}}\right)=\qty{15}{\decibel}\,.
\end{equation}

A primary use case for this monitor is the identification of unshielded oscillators and other narrowband emitters at the \ac{DRAO}. Because this task is analogous to spectral-line observing, we use the \ac{SKA} spectral-line fractional bandwidth to define our system's maximum channel bandwidth. Applying the 0.001\% requirement to the monitor's lowest calibrated frequency of \qty{350}{\mega\hertz} yields a minimum required channel bandwidth of $\frac{0.001}{100}\times\qty{350}{\mega\hertz} = \qty{3.5}{\kilo\hertz}$.

In \S~\ref{sec:sensitivity-validation}, we validate the sensitivity of the commissioned \ac{RFI} monitor against these targets and compare it to commercial off-the-shelf receiver systems.

\subsection{Antenna and Structure} \label{sec:antenna-and-structure}

The requirement established in \S~\ref{sec:antenna-position-req} mandates a minimum path loss of \qty{64}{\decibel} between a transmitting cellular or mobile device on site and the \ac{RFI} monitor. At a cellular uplink frequency of $f=\qty{836.5}{\mega\hertz}$ (wavelength $\lambda=\qty{0.36}{\meter}$), standard \ac{FSPL} requires a physical separation distance of approximately \qty{45}{\meter} to achieve this attenuation.

We chose to mount the monitor antenna on the roof of the main \ac{DRAO} office building, a central location that provides a clear line of sight to most of the site's telescopes. However, because the structurally permissible installation site on this roof is less than \qty{45}{\meter} away from the parking lot and other high-traffic, publicly accessible areas on site, \ac{FSPL} alone is insufficient to prevent receiver saturation. To satisfy the attenuation requirement, we utilize the metal flashing of the building's parapet as a diffracting knife-edge. With the mast's horizontal position locked by these structural limits, the antenna height becomes the sole variable parameter available to optimize the diffraction loss.

\begin{figure}[htb!]
    \centering
    \includegraphics{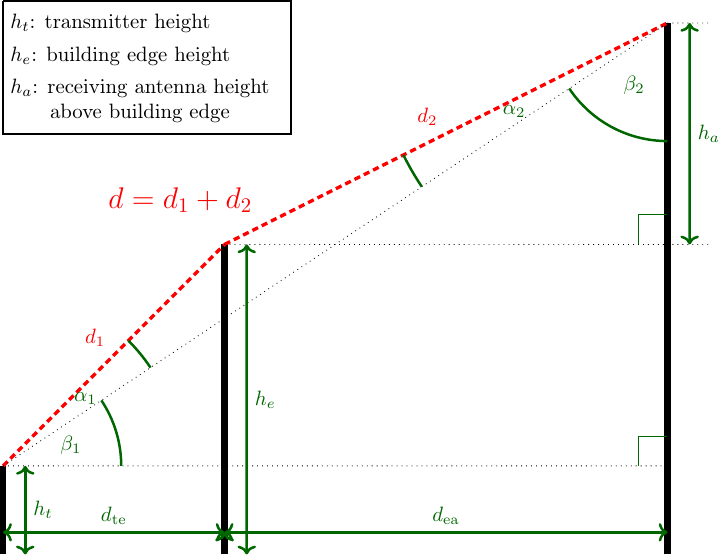}
    \caption{Antenna position geometric model, where $h_t$ is the transmitter height, $h_e$ is the height of the diffracting building edge, and $h_a$ is the receiving antenna height above the building edge.}
    \label{fig:ant-geometric-model}
\end{figure}

To determine the required antenna height, we construct a geometric model (\Cref{fig:ant-geometric-model}) using the following parameters and assumptions:
\begin{itemize}
    \item The parapet flashing acts as a diffracting knife-edge at a height of $h_e = \qty{8.56}{\meter}$ above the ground.
    \item The height of the antenna above the parapet edge, $h_a$, is the variable design parameter (yielding an antenna elevation of $h_e + h_a$).
    \item A structural assessment dictates a fixed horizontal standoff distance from the building edge to the antenna mast of $d_{ea}=\qty{10}{\meter}$.
    \item The mobile transmitter is located at a height of $h_t = \qty{2}{\meter}$, the standard propagation model assumption for a pedestrian or vehicle.
    \item The horizontal distance from the transmitter to the building edge, $d_{te}$, can vary between \qtyrange{1}{100}{\meter}.
    \item Ground reflections are neglected in this geometric model.
\end{itemize}

This geometry is shown in \Cref{fig:ant-geometric-model}. Following the conventions of \citet{itu526}, $d_1$ and $d_2$ represent the slant distances from the diffracting edge to the transmitter and the monitor, respectively. The angles $\alpha_1$ and $\alpha_2$ represent the diffraction angles between the building edge and the respective endpoints.

The total propagation loss, $L$, comprises two distinct components. The first is the \ac{FSPL} defined by \citet{itu525} as
\begin{equation*} \label{eq:loss-fspl}
    20\log_{10}\left(\frac{4\pi d}{\lambda}\right)\,,
\end{equation*}
where $d$ is the total path distance. The second is the additional attenuation due to knife-edge diffraction defined by \citet{itu526} as
\begin{equation*} \label{eq:loss-knife-edge}
    6.9 + 20\log_{10}\left(\sqrt{\left(\nu-0.1\right)^2+1}+\nu-0.1\right)\,,
\end{equation*}
where $\nu = \sqrt{(2d/\lambda)\alpha_1\alpha_2}$ is the dimensionless Fresnel diffraction parameter. Combining these two components yields the total propagation loss constraint:
\begin{equation}\label{eq:ant-propagation-loss}
    L = 20\log_{10}\left(\frac{4\pi d}{\lambda}\right) + 6.9 + 20\log_{10}\left(\sqrt{\left(\nu-0.1\right)^2+1}+\nu-0.1\right)\,.
\end{equation} 
Note that at the grazing angle ($\alpha_1=\alpha_2=0$), the diffraction parameter reduces to $\nu=0$ corresponding to half of the first Fresnel zone being obstructed. Under this condition, the knife-edge introduces $\approx$\qty{6}{\decibel} of additional attenuation. Consequently, the total loss is \qty{6}{\decibel} greater than \ac{FSPL} alone.

\begin{figure}[htb!]
    \centering
    \includegraphics{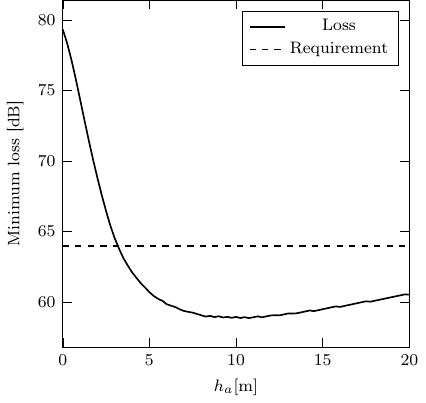}
    \caption{Worst-case path loss evaluated across all transmitter distances ($d_{te}$) as a function of antenna height ($h_a$). To guarantee the target loss of \qty{64}{\decibel}, the antenna must be mounted lower than \qty{4.4}{\meter} above the diffracting edge.}
    \label{fig:ant-geometry-loss-vs-dte}
\end{figure}

Because the transmitter distance ($d_{te}$) is highly variable, we must ensure the \qty{64}{\decibel} attenuation requirement is met under worst-case spatial conditions. For a given antenna height ($h_a$), we calculate the path loss across the entire range of possible pedestrian locations ($d_{te} \in [1, 100]\,\unit{\meter}$) and identify the minimum resulting loss. Plotting the worst-case loss as a function of $h_a$ (\cref{fig:ant-geometry-loss-vs-dte}) demonstrates that the antenna height must be constrained to $h_a\leq\qty{4.4}{\meter}$ to guarantee sufficient attenuation.

\begin{figure}[htb!]
    \centering
    \includegraphics[scale=0.5]{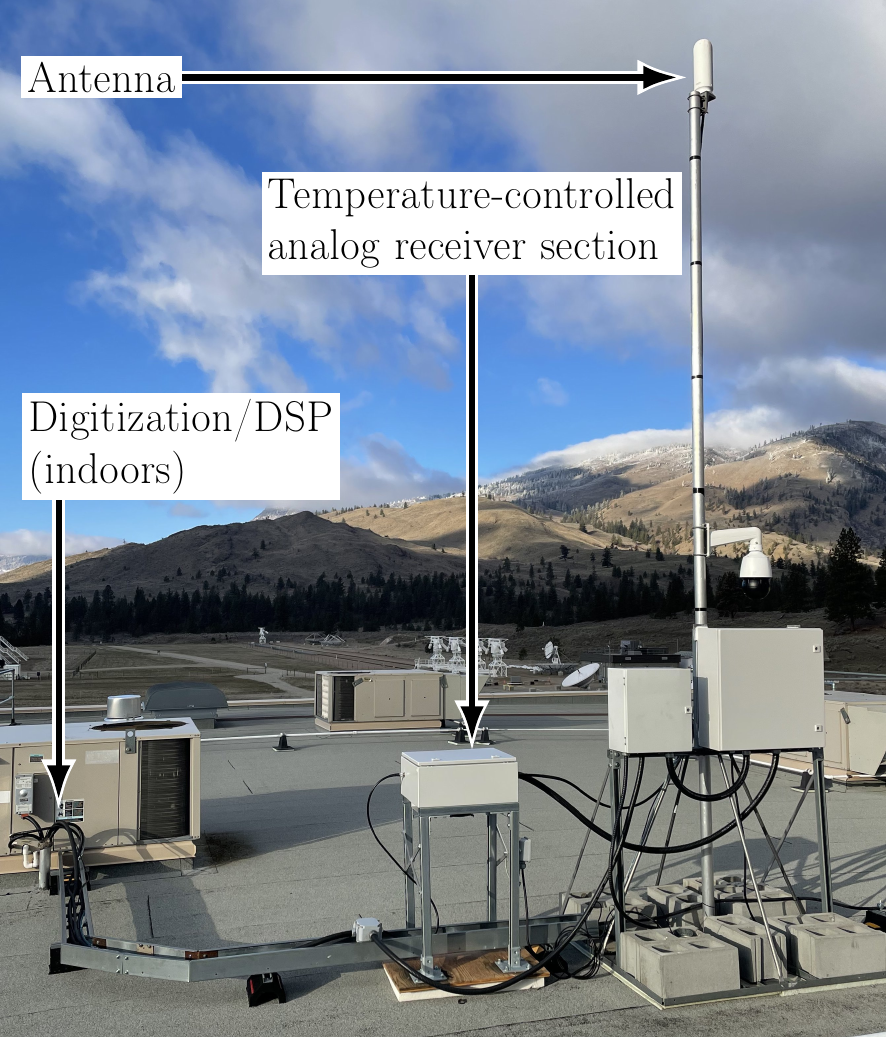}
    \caption{The passive Alaris OMNI-A0190 monitoring antenna mounted on the roof of the main building, alongside the delineated sections of the downstream signal chain. The mounting height is strictly constrained by the knife-edge diffraction loss required to shield the receiver from publicly accessible areas.}
    \label{fig:after-changes-antenna-and-location}
\end{figure}

To provide complete \qty{360}{\degree} coverage of the terrestrial \ac{RF} environment and capture all local interference events, the system requires a vertically polarized, wideband, omnidirectional antenna. We selected the passive Alaris OMNI-A0190 compact monitoring antenna, which operates from \qty{20}{\mega\hertz} to \qty{6}{\giga\hertz}. Across the target frequency range of \qtyrange{350}{1800}{\mega\hertz}, the isotropic antenna gain is nominally \qty{0}{\decibelisotropic} with a minimum of \qty{-4}{\decibelisotropic} at \qty{625}{\mega\hertz} \citep{alarisomni}. The antenna azimuth ripple is $<\qty{6}{\decibel}$ with an $E$-plane beamwidth of \qty{60}{\degree}. The antenna's $E$-plane beamwidth favors the horizon, enabling monitoring of the surrounding area but limiting sensitivity to overhead emitters. Future work may address this through a temporary antenna substitution or a hemispherical replacement, allowing characterization of satellite and aircraft emissions. The installation---comprising the antenna, the mounting structure, and the delineated sections of the signal chain---is shown in \Cref{fig:after-changes-antenna-and-location}.

\subsection{Mechanical Structure} \label{sec:mechanical-structure}

To ensure temperature stability, a copper block of mass $\approx\qty{20}{\kilo\gram}$ serves as a common thermal backplane into which the \ac{RF} signal chain is partially embedded. A direct-to-air \ac{TEA} mediates heat exchange between the copper plate and the external environment while maintaining thermal isolation.

\begin{figure}[htb!]
    \centering
    \includegraphics{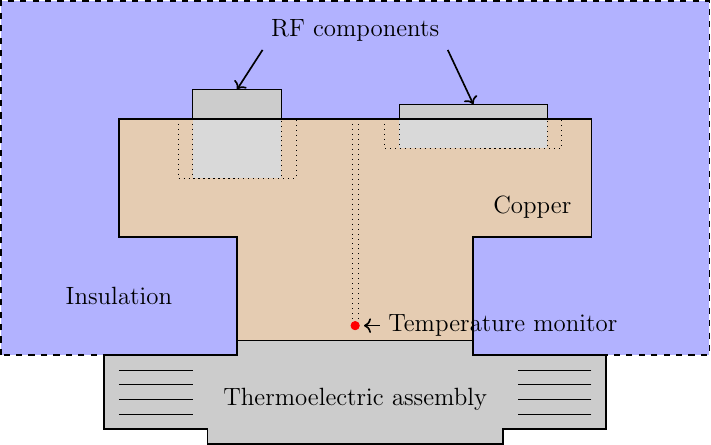}
    \caption{Simplified model of the insulated box containing the \ac{RF} signal chain. Assuming an adiabatic boundary (dashed) and steady-state power dissipation from the \ac{RF} components, heat transfer is restricted exclusively to the \ac{TEA}. Because the feedback sensor is located near the \ac{TEA}--copper interface, ambient temperature variations are actively rejected, and the \ac{RF} components experience negligible temperature variation.}
    \label{fig:adiabatic-box-model}
\end{figure}

A simplified model of the design is shown in \Cref{fig:adiabatic-box-model}. The core design principle is to highly insulate the enclosure to approximate an adiabatic boundary, restricting thermal gradients entirely to the \ac{TEA} interface. By placing the \ac{TEA} control sensor between the \ac{RF} components and the \ac{TEA} itself, the system leverages the thermal mass of the copper. The \ac{TEA} can rapidly compensate for ambient fluctuations before the thermal wave propagates through the block, ensuring the \ac{RF} signal chain remains at a stable temperature. Crucially, there is no requirement for the signal chain to be held at a specific absolute temperature; the sole operational constraint is to limit the temperature-induced gain instability, which is driven by temporal temperature gradients rather than absolute thermal limits.

\begin{figure}[htb!]
    \centering
    \includegraphics{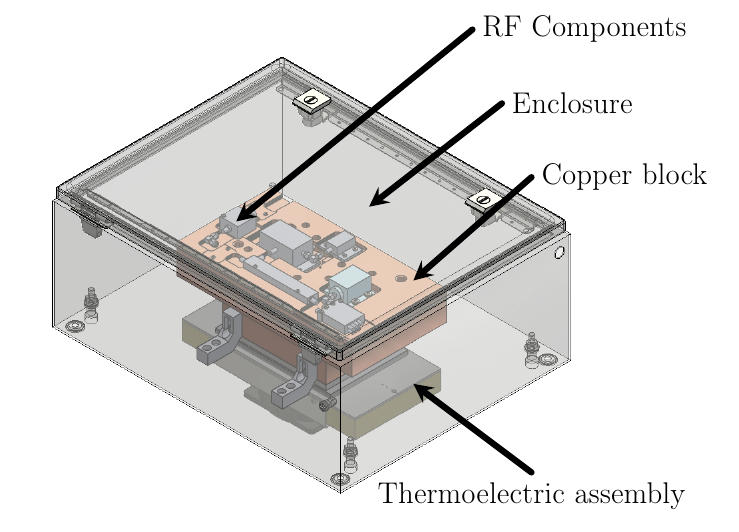}
    \caption{A 3D rendering of the signal chain enclosure illustrating the component layout. To minimize thermal leakage, all depicted void spaces within the physical steel enclosure are completely filled with insulating foam.}
    \label{fig:signal-chain-CAD}
\end{figure}

\begin{figure}[htb!]
    \centering
    \begin{subfigure}{0.44\textwidth}
        \centering
        \includegraphics[width=\textwidth]{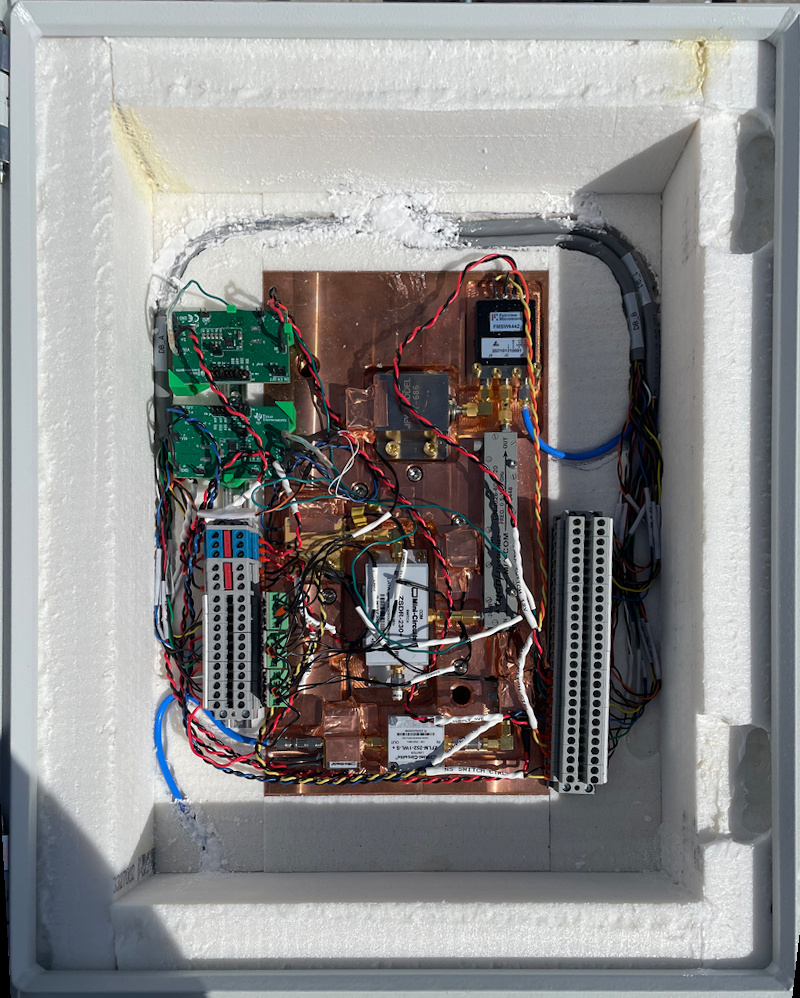}
        \caption{The copper backplane embedded in insulating foam, with signal chain components installed.}
    \end{subfigure}
    \hfill
    \begin{subfigure}{0.44\textwidth}
        \centering
        \includegraphics[width=\textwidth]{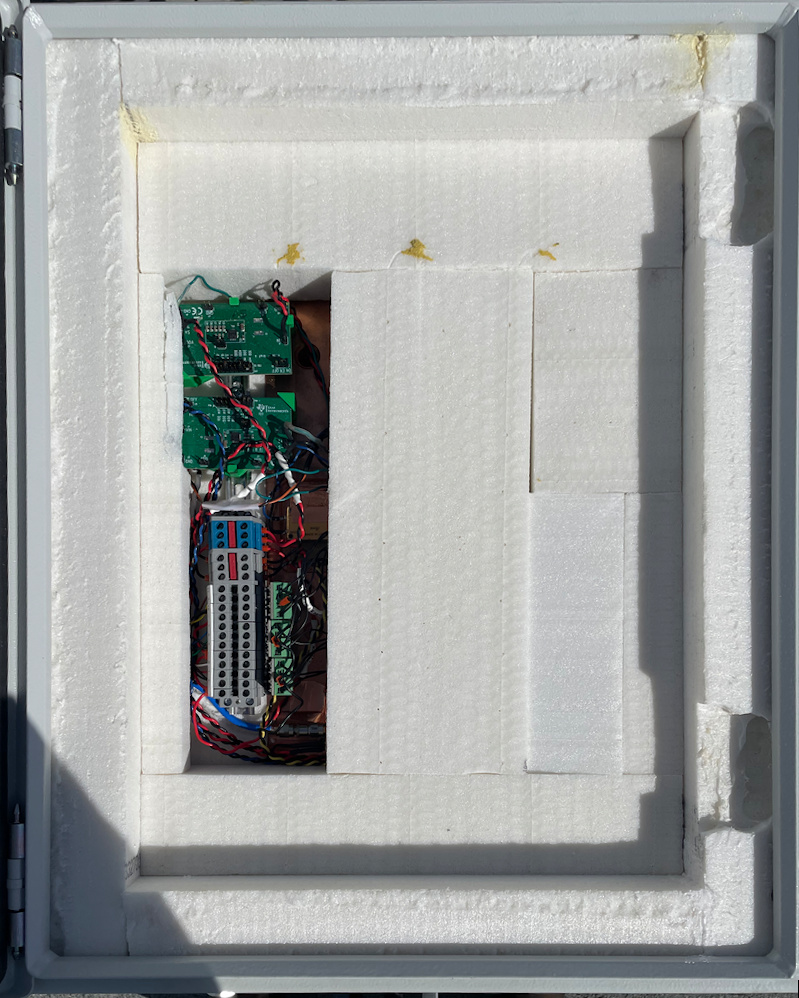}
        \caption{The signal chain components enclosed in precision-cut insulating foam.}
    \end{subfigure}
    \\
    \begin{subfigure}{0.44\textwidth}
        \centering
        \includegraphics[width=\textwidth]{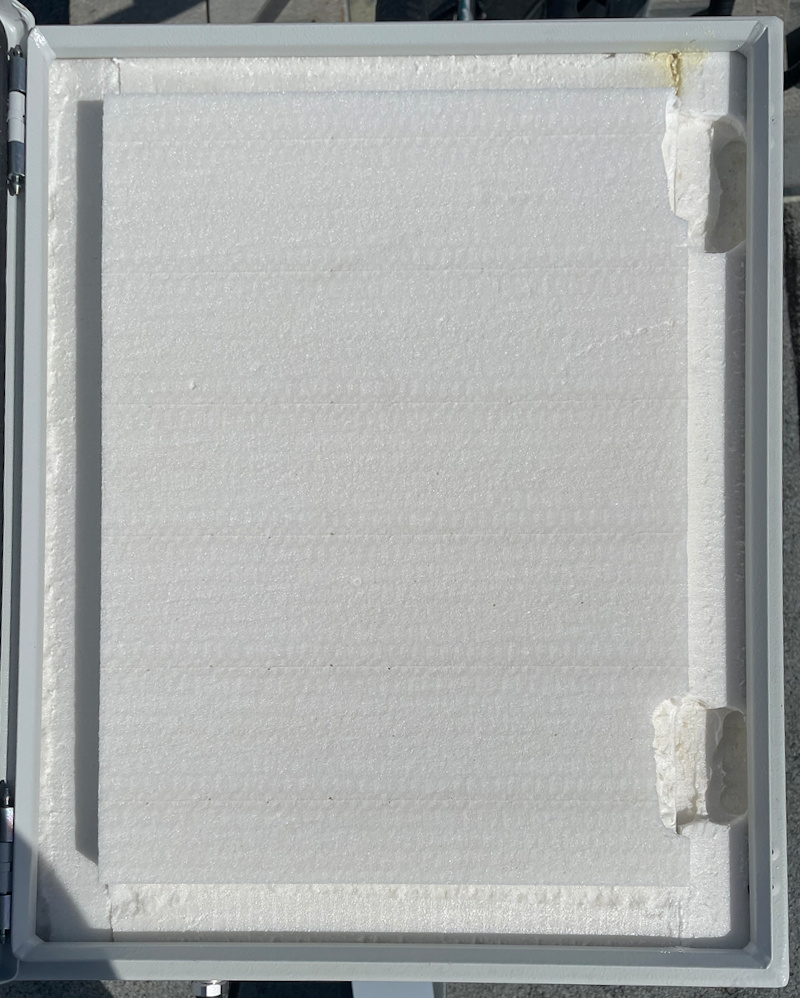}
        \caption{The completed foam insert, fully enclosing the thermally stabilized signal chain.}
    \end{subfigure}
    \caption{The signal chain enclosure at successive stages of foam insulation assembly.}
    \label{fig:signal-chain-insulation}
\end{figure}

A 3D rendering of the signal chain enclosure is shown in \Cref{fig:signal-chain-CAD}. The copper backplane provides a thermal standoff approximately \qty{10.2}{\centi\meter} thick between the \ac{TEA} interface and the embedded \ac{RF} components. To enforce the adiabatic assumption, the environmentally sealed steel enclosure is completely back-filled with rigid, closed-cell insulating foam (Airex T92.60). \Cref{fig:signal-chain-insulation} shows the signal chain and copper mass surrounded by the insulating foam. Wires are embedded into the rigid foam, and spray-foam insulation is used to prevent air gaps.

\subsection{Calibration System} \label{sec:calibration-system}

\begin{figure}[htb!]
    \centering
    \includegraphics[width=0.4\textwidth]{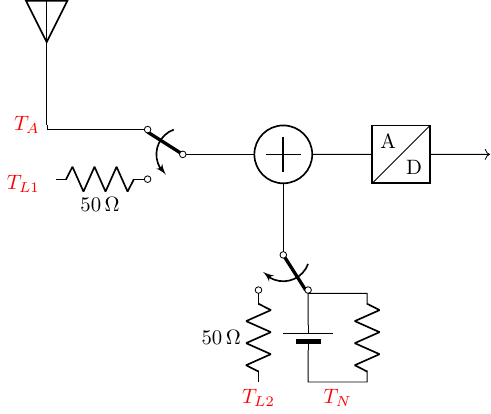}
    \caption{Simplified block diagram of the calibration system architecture. The first switch selects between the antenna ($T_A$) and a primary \qty{50}{\ohm} matched load ($T_{L1}$). The second switch toggles the calibration input between a noise diode ($T_N$) and a secondary \qty{50}{\ohm} matched reference load ($T_{L2}$).}
    \label{fig:calibration-block-diagram}
\end{figure}

To maintain the gain stability of the analog signal chain, the monitor employs a standard $Y$-factor calibration. This process utilizes two discrete states to characterize the system noise: a ``hot'' measurement and a ``cold'' measurement. A simplified block diagram of this architecture is shown in \Cref{fig:calibration-block-diagram}.

Two high-isolation switches control the signal routing for these measurements. During calibration, the first switch disconnects the antenna from the signal path and replaces it with the primary \qty{50}{\ohm} matched load. A ``hot'' measurement is then recorded by toggling the second switch to couple the noise diode into the signal path. Conversely, a ``cold'' measurement is obtained by toggling the second switch to replace the noise diode with a secondary \qty{50}{\ohm} matched load. This dual-switch configuration ensures the receiver remains isolated from the external \ac{RF} input while providing a stable reference for the $Y$-factor calculation. The physical layout and placement of these switching and calibration components within the analog receiver are detailed in \S~\ref{sec:design-analog-receiver}.

\subsubsection{Generating Calibrated Integrations}

The ``hot'' and ``cold'' measurements yield the uncalibrated power spectra, $P_{\rm hot}$ and $P_{\rm cold}$, respectively, in dimensionless \ac{ADC} units. Combined with the known physical temperature of the primary matched load, $T_{L1}$, and the excess noise temperature of the noise diode, $T_N$, these spectra allow us to derive the receiver noise temperature, $T_{\rm rec}$, and the system gain, $G$. 

As illustrated in \Cref{fig:calibration-block-diagram}, the total equivalent noise temperatures for the two states are $T_{\rm hot}=T_{L1}+T_N$ and $T_{\rm cold}=T_{L1}$. We omit the contribution of $T_{L2}$ to $T_{\rm cold}$; because it is injected via a \qty{-20}{\decibel} coupler (detailed in \S~\ref{sec:design-analog-receiver}), its effective noise contribution is on the order of only \qty{1}{\kelvin}. The ratio of the measured powers defines the standard $Y$-factor:
\begin{equation}\label{eq:Y-factor-basic}
    Y = \frac{P_{\rm hot}}{P_{\rm cold}} = \frac{T_{\rm hot}+T_{\rm rec}}{T_{\rm cold}+T_{\rm rec}}\,.
\end{equation}
Solving \cref{eq:Y-factor-basic} for the receiver temperature yields:
\begin{equation} \label{eq:receiver-temp}
    T_{\rm rec} = \frac{T_N}{Y-1} - T_{L1}\,.
\end{equation}

With $T_{\rm rec}$ established, we use the Johnson-Nyquist noise power relation with a gain term, $P=GkT\Delta f$, to determine the system gain, $G$. This factor maps the physical power to the digitized \ac{ADC} output. Evaluating the ``cold'' state ($P=P_{\rm cold}$ and $T=T_{\rm cold}$), the gain is
\begin{equation} \label{eq:calibration-gain}
G = \frac{P_{\rm cold}}{k (T_{L1}+T_{\rm rec}) \Delta f}\,.
\end{equation}
The units of $G$ are \ac{ADC} units per watt. Note that we could equivalently compute the gain using the ``hot'' state parameters ($P_{\rm hot}$ and $T_{\rm hot}$).

This calibration gain is then applied to each uncalibrated integration, $P_{\rm uncal}$ (\ac{ADC} units), to yield a calibrated \ac{PSD} in physical units ($\unit{\watt\per\hertz}$):
\begin{equation} \label{eq:uncalibrated-power}
    S_{x,\rm cal} = \frac{P_{\rm uncal}}{G\Delta f}\,.
\end{equation}
The factor $1/G$ reverses the instrumental gain, converting the raw digitized \ac{ADC} units back into absolute power. Finally, $S_{x,\rm cal}$ is converted to logarithmic units (\unit{\decibelmilliwatt\per\hertz}) to allow for direct comparison against the \ac{RFI} sensitivity requirements established in \S~\ref{sec:sensitivity-design}.

\subsubsection{Meeting the On-Sky Requirement}

To maximize the system's sensitivity to transient \ac{RFI}, the monitor must maintain an on-sky observing efficiency of at least 95\%. Satisfying this requirement necessitates a careful balance: the calibration cycle must be as brief as possible, yet long enough to achieve an acceptable signal-to-noise ratio in the $Y$-factor measurement.

To determine the minimum number of averaged integrations, $N$, required for calibration, we assess the uncertainty in the $Y$-factor. Assuming the noise power in each frequency channel follows a normal distribution, the standard deviation of an averaged measurement decreases by a factor of $1/\sqrt{N}$. Let $\sigma_{\rm hot}$ and $\sigma_{\rm cold}$ represent the standard deviations of a \textit{single} integration for the hot and cold states, respectively. If we average $N_{\rm hot}$ and $N_{\rm cold}$ integrations to form the mean power measurements $P_{\rm hot}$ and $P_{\rm cold}$, respectively, the standard errors of these means are $\sigma_{\rm hot}/\sqrt{N_{\rm hot}}$ and $\sigma_{\rm cold}/\sqrt{N_{\rm cold}}$.

Applying the standard formula for propagation of uncorrelated errors to the $Y$-factor quotient ($Y = P_{\rm hot}/P_{\rm cold}$), we obtain the variance:
\begin{equation}
    \left(\frac{\sigma_Y}{Y}\right)^2 = \left(\frac{\sigma_{\rm hot}}{P_{\rm hot}\sqrt{N_{\rm hot}}}\right)^2 + \left(\frac{\sigma_{\rm cold}}{P_{\rm cold}\sqrt{N_{\rm cold}}}\right)^2\,,
\end{equation}
which simplifies to the final form used to evaluate the calibration duration:
\begin{equation}
    \frac{\sigma_Y}{Y} = \sqrt{\frac{\sigma_{\rm hot}^2}{N_{\rm hot}P_{\rm hot}^2}+\frac{\sigma_{\rm cold}^2}{N_{\rm cold}P_{\rm cold}^2}}\,.
\end{equation}

To empirically establish these parameters, we recorded 500 independent integrations of both the hot and cold calibration states (as defined in \S~\ref{sec:calibration-system}). Using the sample means and standard deviations across all frequency channels, we calculated the fractional $Y$-factor uncertainty for symmetric integration counts ($N=N_{\rm hot}=N_{\rm cold}$).

\begin{figure}[htb!]
    \centering
    \includegraphics{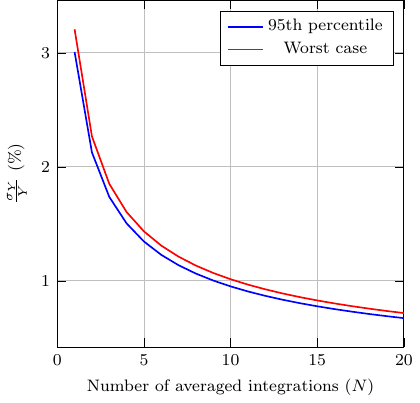}
    \caption{Fractional $Y$-factor uncertainty as a function of the number of averaged integrations ($N$). The curves represent the maximum (worst-case; red) and 95th percentile (blue) values across the measured spectrum. At $N=9$, 95\% of the frequency bins achieve a fractional $1\sigma$ uncertainty of less than 1\%.}
    \label{fig:fractional-Y-factor-uncertainty}
\end{figure}

As shown in \Cref{fig:fractional-Y-factor-uncertainty}, evaluating $\sigma_Y/Y$ across all frequencies demonstrates that averaging $N=9$ integrations ensures the $1\sigma$ uncertainty remains below 1\% for 95\% of the frequency bins. With an integration time of \qty{0.75}{\second}, recording 9 integrations for both the hot and cold states results in a total calibration duration of \qty{13.5}{\second}.

To ensure this \qty{13.5}{\second} calibration overhead consumes no more than 5\% of the monitor's operational time, the system must execute a maximum of 320 calibration cycles per day. This establishes a strict calibration cadence (the duration between the start of successive calibrations) of exactly \qty{270}{\second}. Within this cadence, \qty{13.5}{\second} are spent calibrating, leaving \qty{256.5}{\second} for active \ac{RF} environment monitoring.

To guarantee this precise \qty{270}{\second} cadence without being subject to the unpredictable latency of software-level execution, the analog signal chain switches are controlled directly by the \ac{ADC}'s \ac{GPIO} pins. The \ac{FPGA} implements a hardware-level sequencer; pairs of timestamps and control values are pushed from the software into a \ac{FIFO} block. As the internal system clock progresses, entries are dequeued from the \ac{FIFO} and applied directly to the \ac{GPIO} pins, ensuring microsecond-level switching alignment with the digitized samples. The exact state of the switches is permanently embedded within the \ac{VITA} 49.0 metadata trailer of each integration, allowing any actuation delays to be compensated downstream and guaranteeing seamless application of the calibration coefficients.

\subsection{Analog Receiver} \label{sec:design-analog-receiver}

\begin{figure}[htb!]
    \centering
    \includegraphics[scale=0.75]{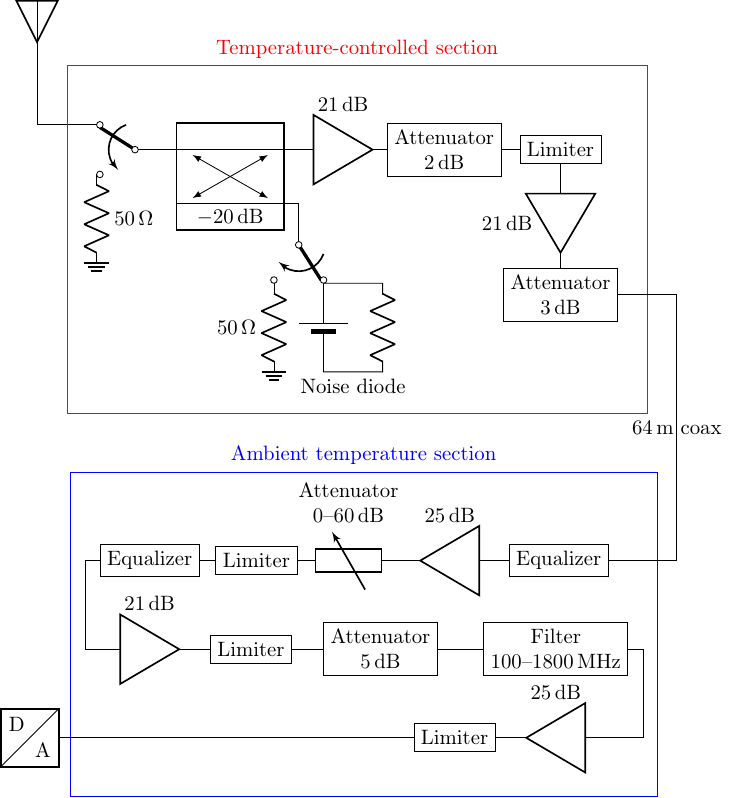}
    \caption{Analog receiver block diagram. The temperature-controlled section is deployed outdoors adjacent to the antenna mast and contains the calibration hardware. The ambient section is deployed indoors and contains an equalizer, adjustable gain, and an anti-aliasing filter. The sections are connected by \qty{64}{\meter} of coaxial cable.}
    \label{fig:analog-section-block-diagram}
\end{figure}

A simplified block diagram of the analog receiver is shown in \Cref{fig:analog-section-block-diagram}. The signal chain is divided into two distinct physical environments: a temperature-controlled front end deployed outdoors adjacent to the antenna mast (\Cref{fig:after-changes-antenna-and-location}), and an ambient-temperature section collocated indoors with the \ac{DSP} subsystem. 

The temperature-controlled section contains the calibration hardware and the first-stage \acp{LNA} to minimize front-end insertion loss and ensure optimal gain stability. The primary \qty{50}{\ohm} matched load (JFW 50T-686) is bolted directly into the copper backplane (see \S~\ref{sec:mechanical-structure}) to leverage its immense thermal mass. A Fairview FMSW6441 latching switch routes the signal path between the antenna and this matched load, utilizing two \acs{TTL} inputs to direct the mechanical latch to either \ac{RF} port. Because the switch requires no continuous holding current, it dissipates negligible heat, preserving the system's temperature stability during calibration. Furthermore, its low insertion loss of \qty{0.3}{\decibel} minimizes its contribution to the system noise temperature.

\begin{figure}[htb!]
    \centering
    \includegraphics{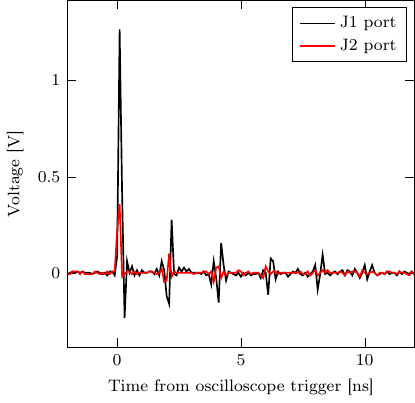}
    \caption{Time-domain transients generated by the latching switch at the antenna output. The J1 port exhibits a more energetic response; consequently, J2 is utilized for the antenna. The subsequent reflection pulses are an artifact of a coaxial mismatch in the experimental setup and do not occur in the installed system.}
    \label{fig:latching-switch-transients}
\end{figure}

Mechanical latching switches intrinsically emit electrical transients during actuation. To characterize this effect, high-resolution time-domain samples were captured from each \ac{RF} port immediately after actuation using a \qty{20}{\giga\samplepersecond} bandwidth oscilloscope (\Cref{fig:latching-switch-transients}). Testing revealed that the J1 port exhibits a significantly higher-energy transient; consequently, the quieter J2 port was selected for the antenna path. Crucially, measurements confirmed that the transient is exclusively visible on the active port, indicating no measurable leakage to the disconnected path. However, any transient emitted from the active port will couple directly into the antenna and broadcast as localized \ac{RFI}. While a solid-state switch (SKY13373-460LF) was evaluated and found to produce a smaller transient, it exhibited higher insertion loss and inferior port isolation compared to the mechanical switch. The mechanical switch was ultimately selected because its transient duration is exceedingly brief ($<\qty{1}{\nano\second}$) and occurs at a predictable, regular cadence, making it straightforward to identify and flag in the local telescope data. The secondary reflection pulses observed on the oscilloscope were caused by a coaxial mismatch in the experimental setup. These subsequent pulses are not present in the installed system. 

The signal travels indoors via \qty{64}{\meter} of coaxial cable to the ambient-temperature section. This stage equalizes the line loss, applies programmable attenuation, boosts the signal for the \ac{ADC}, and provides low-pass anti-aliasing filtering (\qtyrange{100}{1800}{\mega\hertz}). To manage the system gain dynamically, the control software interfaces directly with networked programmable attenuators, allowing attenuation levels to be adjusted remotely during periods of elevated \ac{RFI} (such as during facility weekend tours).

To protect the amplifiers and the \ac{ADC} from damage or saturation during severe \ac{RFI} events, three signal limiters are distributed throughout the signal chain. The first two (Mini-Circuits ZFLM-252-1WL-S+) are used to protect the subsequent amplifiers; they feature a \qty{0.1}{\decibel} compression point at \qty{-10}{\decibelmilliwatt} and a maximum output power of \qty{0}{\decibelmilliwatt}. The final limiter (Mini-Circuits VLM-33S+) protects the \ac{ADC}. It has a maximum output power of \qty{12}{\decibelmilliwatt}, and at an input power of \qty{1}{\decibelmilliwatt}, its \ac{SFDR} is approximately \qty{60}{\decibelcarrier}. For comparison, the \ac{ADC} saturates at \qty{0.96}{\decibelmilliwatt}, possesses a maximum absolute input limit of \qty{17}{\decibelmilliwatt}, and has an \ac{SFDR} of approximately \qty{50}{\decibelcarrier}. The limiters provide sufficient dynamic range and linearity without introducing spurious intermodulation artifacts into the data.

\subsection{Digital Signal Processing}

\begin{figure}[htb!]
    \centering
    \includegraphics[scale=0.75]{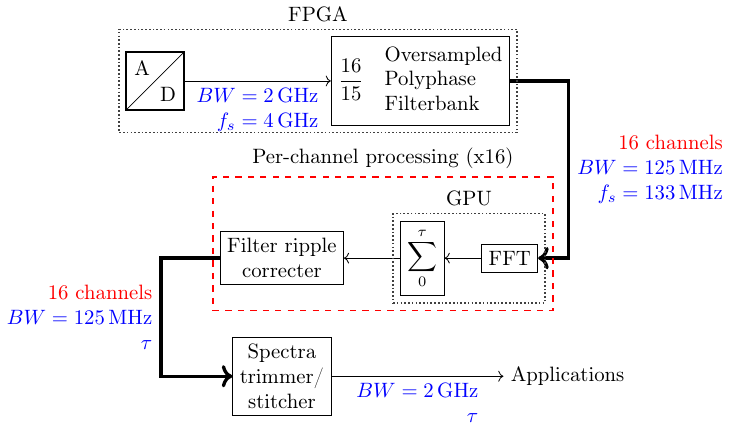}
    \caption{\ac{DSP} block diagram detailing the 16-channel oversampled filterbank implemented in the \ac{FPGA}, which feeds into 16 critically sampled channelizers implemented in the \acs{GPU}.}
    \label{fig:dsp-block-diagram}
\end{figure}

The \ac{DSP} subsystem utilizes an architecture typical of a single-dish backend and largely leverages prior design work developed for other telescopes at the \ac{DRAO}. This includes precise timing controls for the calibration noise source in the analog receiver, ensuring strict synchronization with the integration time of the spectrometer. \Cref{fig:dsp-block-diagram} illustrates the data flow through the \ac{DSP} subsystem, which interfaces directly with the output of the analog signal chain (\Cref{fig:analog-section-block-diagram}). The \ac{RF} signal is digitized into complex (I/Q) samples using an interleaved \ac{ADC}, such that a \qty{4}{\giga\samplepersecond} sampling rate yields the full \qty{2}{\giga\hertz} of usable bandwidth. 

The first \ac{DSP} stage is \iac{FPGA}-based \ac{OSPFB} (oversampling factor $O=16/15$) that produces 16 single-polarization channels from the digitized input. The sample rate for each channel is \qty{133}{\mega\hertz}, providing a usable Nyquist bandwidth of \qty{125}{\mega\hertz}. This architecture renders each of the 16 channels comparable to the bandwidth of a high-end \ac{SDR}, such as the Ettus \ac{USRP} X310. Although the channel center frequencies are fixed, the filterbank provides contiguous, full-frequency coverage over the entire range of interest. 

\begin{figure}[htb!]
   \centering
    \includegraphics[scale=0.75]{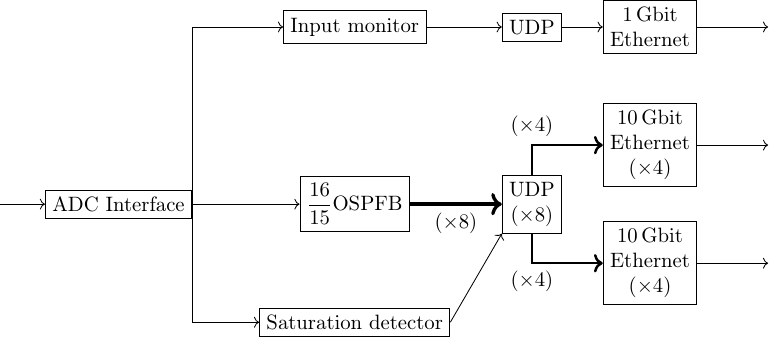}
    \caption{\ac{FPGA} \ac{DSP} block diagram showing the digitized data being processed by the \ac{OSPFB}, a saturation detector, and an input monitor, all of which broadcast their outputs over \ac{UDP} channels.}
   \label{fig:fpga}
\end{figure}

The \ac{OSPFB} is implemented on the ICE platform \citep{bandura2016ice} and relies on its native board management and control software. The internal architecture of these \ac{FPGA} \ac{DSP} stages is shown in \Cref{fig:fpga}. The design directly reuses code from the \ac{CHIME} project, as well as legacy firmware from previous instruments based on the Kermode \ac{FPGA} platform \citep{harrison2018software, harrison2018digital}. The ICE platform streams data to downstream compute resources via two \qty{10}{\giga\bit} Ethernet links. Each link carries eight single-polarization channels, packetized as \qty{16}{\bit}+\qty{16}{\bit} complex samples and framed according to the \ac{VITA} 49.0 radio transport standard \citep{vita49}. Each \ac{UDP} packet contains two single-polarization channels. Each \qty{10}{\giga\bit} link publishes four streams that are separated into eight single-polarization channels by the second \ac{DSP} stage. Crucially, all bit growth is preserved from the \ac{OSPFB} through to the second-stage \ac{GPU} processing, employing a similar approach to that of \citet{hobbs2020}.

The second \ac{DSP} stage is \iac{GPU}-based, critically sampled fine channelizer hosted on a high-performance server. The server ingests the raw time-domain packets from the two Ethernet links using \ac{DPDK} and routes the data directly to a bank of two NVIDIA GeForce RTX 2080 Ti \acp{GPU} where channelization and accumulation occur. Following accumulation, the pipeline mathematically corrects the filter ripple introduced by the first-stage \ac{OSPFB}.

For each integration cycle, the resulting data products are published via 16 \ac{ZeroMQ} streams, which are available to any subscribed service on the local network. By design, the spectrometer's integration time and frequency channel bandwidth are interdependent parameters. For example, achieving a channel bandwidth of \qty{3.33}{\kilo\hertz} dictates a minimum integration time of $\approx$ \qty{50}{\milli\second} (constrained by memory bandwidth), whereas an integration time of \qty{1}{\second} allows for a minimum channel bandwidth of $\approx$ \qty{100}{\hertz} (constrained by \ac{GPU} memory).

For standard operations, a channel bandwidth of \qty{3.33}{\kilo\hertz} was selected to meet the specific \ac{RFI} requirements defined in \S~\ref{sec:requirements}. Concurrently, an integration time of \qty{0.75}{\second} was chosen to manage data storage volumes. Note that these internal hardware integration times are distinct from the $\tau$-duration integrations discussed in \S~\ref{sec:gain-stability-requirement}. Because the raw spectra are output at a fast \qty{0.75}{\second} cadence, downstream applications can subsequently average these data to achieve an effective integration time of $\tau$, or any longer duration demanded by specific science goals.

These 16 data streams can be individually subscribed to by various software applications. The primary downstream application subscribes to all 16 streams, trims the overlapping oversampled channel edges, concatenates the data, and republishes a single, unified \ac{ZeroMQ} stream containing contiguous \qty{3.33}{\kilo\hertz} channels from \qtyrange{0}{2}{\giga\hertz} with unified metadata. This main stream provides the definitive data product used by the majority of the facility's monitoring services.

\subsection{Observing Software}

The ``publishing pipeline'' continues in the observing software when a calibrated integration publisher subscribes to the concatenated, uncalibrated spectra, applies a $Y$-factor calibration, and republishes each integration. This architecture allows downstream software to subscribe to either calibrated or uncalibrated spectra and receive a full-bandwidth data product for every integration.

The fundamental data products generated by the system are one calibrated and one uncalibrated file every hour, each containing exactly one hour of data. These files are saved in \ac{SigMF}, which has been modified for use with power spectra. These files are stored locally for two weeks, during which time they are available for download and can be explored using a web-based tool. This tool renders a selected hour of data as a time-frequency ``waterfall'' plot that can be interactively panned and scaled at full resolution. After two weeks, a reduced statistical dataset is automatically archived, and the full-resolution data files are deleted.

A primary obstacle to the success of previous \ac{RFI} monitoring projects at the \ac{DRAO} has been the sheer volume of data produced. Massive datasets are costly to store and often yield limited immediate insight into the local \ac{RF} environment. To address this, we employ the data collection and analysis approach previously presented in \citet{harrison2019}, which represents a large number of events in a compact, easily mineable format. Using this method, individual signals are autonomously detected, and key features---such as duration, bandwidth, power, and time of day---are stored in place of the raw spectra. Future integration of the analysis technique presented in \citet{bruce_nicholas_ddc_2022} will enable baseband signal reconstruction. Subsequent analysis of these baseband signal representations, using methods such as the unsupervised modulation recognition introduced in \citet{bruce_nicholas_modrec_2024}, will further expand the feature set for each detection. 

\subsection{Instrument Software}

All of the system software is containerized, and the monitoring and control of the instrument are handled by a Tango Controls system \citep{tango}. Most telemetry data consist of temperatures collected by the LabJack T4, which are sent to a time-series database. This database is connected to a web front-end that allows users to easily visualize the telemetry data and configure temperature alerts.

\begin{figure}[htb!]
    \centering
    \includegraphics{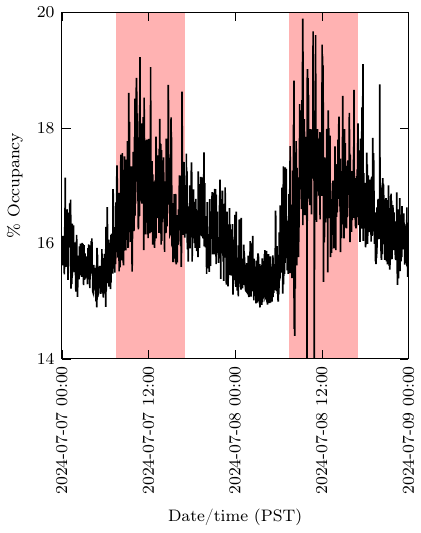}
    \caption{Two days of occupancy data for the \qtyrange{350}{1050}{\mega\hertz} band, demonstrating a strong diurnal cycle. Standard working hours (07:30--17:30) are shaded in red.}
    \label{fig:watched-bands}
\end{figure}

A dedicated microservice monitors 23 specific ``bands of interest.'' These encompass protected radio astronomy bands focused on single frequencies (e.g., \qtylist{408;1420}{\mega\hertz}), active local cellular bands (e.g., \qtyrange{824}{829}{\mega\hertz}), and broad operational bandwidths for on-site telescopes (such as \qtyrange{400}{800}{\mega\hertz} for \ac{CHIME}). The service subscribes to the calibrated integrations and, depending on the band, computes a subset of metrics for each integration, including the mean power, maximum power, spectral occupancy, flux, and centroid. These data are also sent to the time-series database and have provided valuable insights into local cellular activity. \Cref{fig:watched-bands} shows a diurnal cycle of occupancy for one such band of interest. This variation is likely attributable to the routine influx of delivery drivers, employees, and contractors to the site during standard working hours.

\section{Validation} \label{sec:validation}

\subsection{Receiver Noise Temperature} \label{sec:Trec-validation}

\begin{figure}[htb!]
    \centering
    \includegraphics{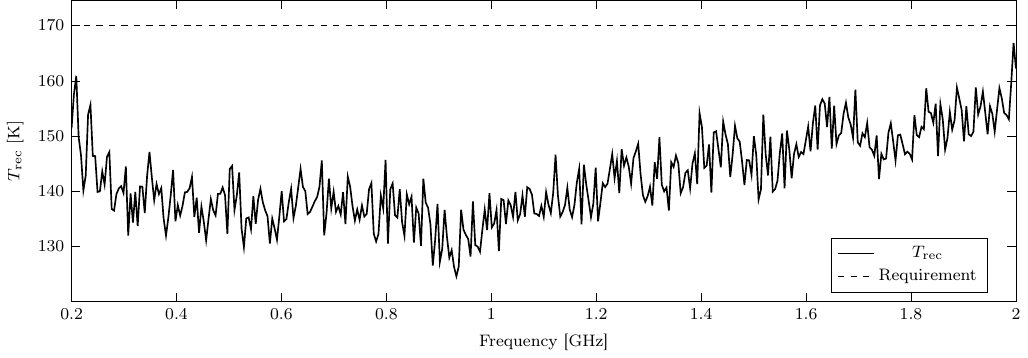}
    \caption{Receiver noise temperature across the operational bandwidth, demonstrating a maximum of less than \qty{170}{\kelvin}.}
    \label{fig:Trec}
\end{figure}

To determine the receiver noise temperature, $T_{\rm rec}$, we measured the analog signal chain using \iac{NFA}. \Cref{fig:Trec} demonstrates that the requirement of $T_{\rm rec}<\qty{170}{\kelvin}$, established in \S~\ref{sec:sys-temp-requirement}, is successfully met across the entire bandwidth. An approximately \qty{5}{\meter} length of LMR-600 coaxial cable connects the antenna output to the analog signal chain input, contributing $\approx$\qty{40}{\kelvin} to the overall system noise temperature. To validate the physical measurement, we computed a comprehensive \ac{RF} link budget, which predicts $T_{\rm rec}=\qty{134.7}{\kelvin}$ at \qty{1}{\giga\hertz}. This theoretical prediction aligns exceptionally well with the empirical measurement of $T_{\rm rec}=\qty{133.5}{\kelvin}$ at \qty{1}{\giga\hertz} shown in \Cref{fig:Trec}.

\subsection{Temperature-Induced Gain Stability} \label{sec:gain-stability}

To characterize the temperature-induced gain instability described in \S~\ref{sec:gain-stability-requirement}, we used an environmental chamber to sweep the ambient temperature while \iac{VNA} recorded the corresponding gain variations.

\begin{figure}[htb!]
    \centering
    \includegraphics{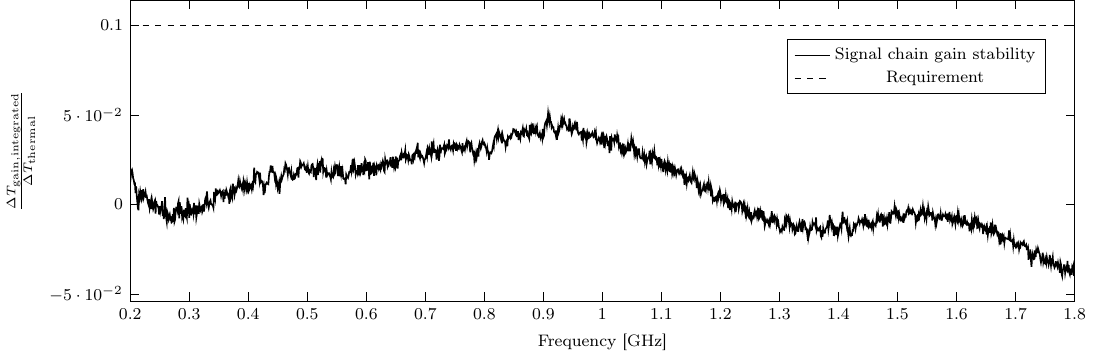}
    \caption{Gain stability of the system, demonstrating that the performance requirement is safely met over a \qty{1200}{\second} duration across the entire operational bandwidth.}
    \label{fig:gain-stability-requirement}
\end{figure}

For the \qty{1200}{\second} comparison period defined in \cref{eq:integrated-gain-uncertainty-requirement}, the most extreme temperature change recorded at the \ac{DRAO} in 2024 was \qty{3.3}{\celsius} (from \qtyrange{26.0}{29.3}{\celsius}). Our objective was to verify that the system meets the stability requirement under these extreme conditions, which correspond to a constant temperature ramp rate of \qty{0.165}{\celsius\per\minute}. We cycled the environmental chamber across this range at this precise rate while the \ac{VNA} continuously recorded the system's gain. \Cref{fig:gain-stability-requirement} compares these empirical measurements against the theoretical upper bound established in \cref{eq:integrated-gain-uncertainty-requirement}.

\subsection{Integrated System Stability}

The final validation test was a repeated characterization of the system's Allan deviation, following the methodology described in \S~\ref{sec:gain-stability-requirement}. The commissioned system was reinstalled at the same physical location on the \ac{DRAO} main building roof where the prototype had been evaluated. Data were collected over a \qty{119}{\hour} period, with the input signal coming from the internal matched load rather than the antenna. The noise diode was switched in for \qty{6.75}{\second} every \qty{300}{\second}. This dataset was utilized to generate two distinct Allan deviation curves: an uncalibrated time series to determine the necessary calibration cadence, and a calibrated time series to demonstrate that long-term integration successfully reduces the noise variance.

\begin{figure}[htb!]
    \centering
    \includegraphics{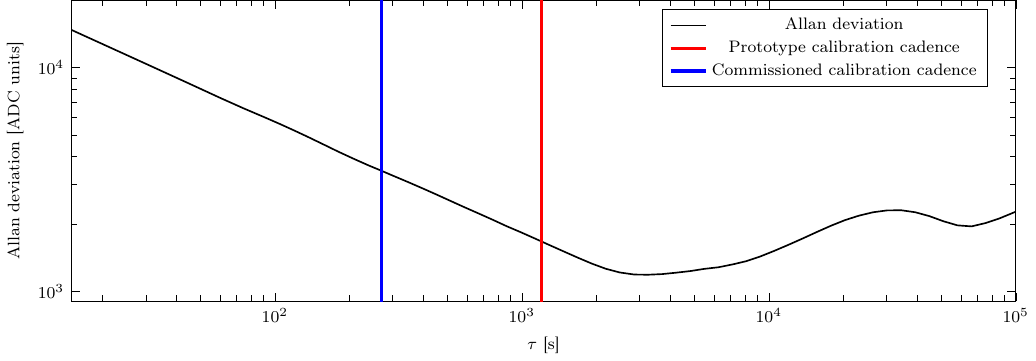}
    \caption{Uncalibrated Allan deviation (black) at \qty{1250}{\mega\hertz}. The \ac{RFI} monitor data cannot be integrated for longer than $\approx\qty{1400}{\second}$ before system drift causes the noise variance to increase. The original prototype calibration cadence of \qty{1200}{\second} (red) occurs prior to this stability knee, and the minimum possible calibration cadence for the commissioned instrument (\qty{270}{\second}, blue) is well to the left of the drift-limited regime.}
    \label{fig:after-changes-uncal-allan-deviation}
\end{figure}

\Cref{fig:after-changes-uncal-allan-deviation} shows the Allan deviation of the uncalibrated system. These results indicate that the system remains stable for approximately twice as long as the prototype. It is now possible to calibrate on the original \qty{1200}{\second} cadence while continuously reducing the noise variance in the measurement. 

\begin{figure}[htb!]
    \centering
    \includegraphics{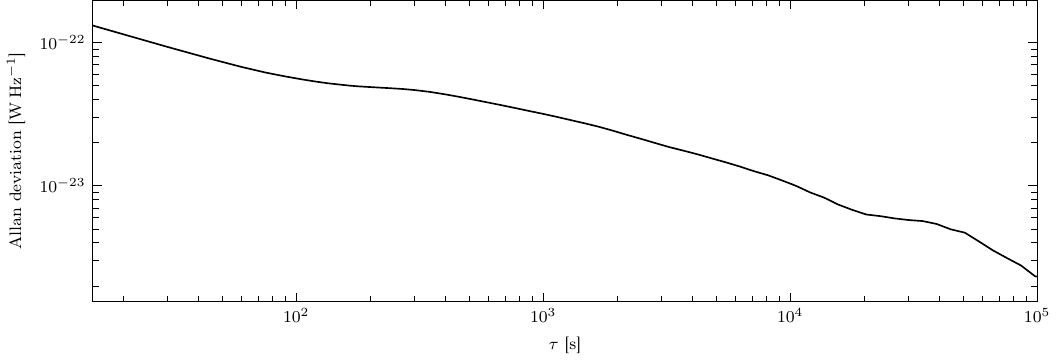}
    \caption{Calibrated Allan deviation at \qty{1250}{\mega\hertz}, demonstrating that the \ac{RFI} monitor data can be integrated up to $\approx\qty{100000}{\second}$ without system drift dominating the noise variance.}
    \label{fig:after-changes-cal-allan-deviation}
\end{figure}

The Allan deviation of the calibrated system is shown in \Cref{fig:after-changes-cal-allan-deviation}. These data demonstrate that the signal can be continuously integrated for at least \qty{100000}{\second} ($\approx$\qty{28}{\hour}) while the noise variance continues to decline. Because no local minimum is yet apparent in \Cref{fig:after-changes-cal-allan-deviation}, no definitive limit for the maximum integration duration can be established. However, we consider a guaranteed \qty{24}{\hour} integration period to be more than sufficient for the \ac{RF} site characterization tasks envisioned in \S~\ref{sec:introduction}.

\subsection{Sensitivity} \label{sec:sensitivity-validation}

\begin{figure}[htb!]
    \centering
    \includegraphics{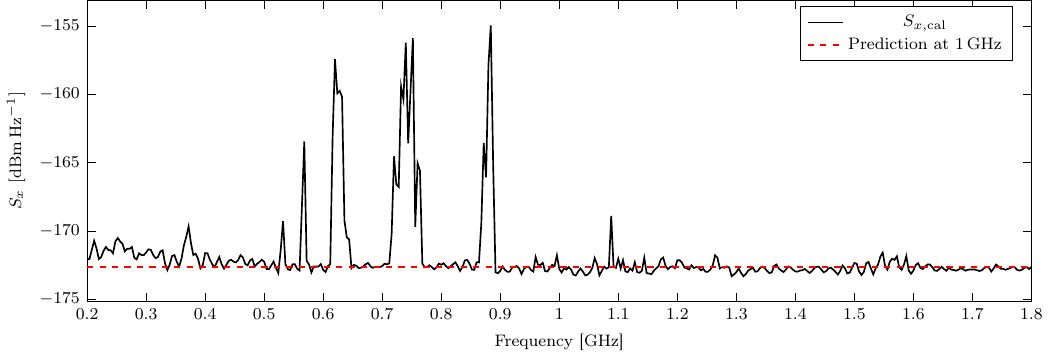}
    \caption{Sample spectra output from the \ac{RFI} monitor showing a measured system sensitivity of $\approx\qty{-172}{\decibelmilliwatt\per\hertz}$, aligning with the theoretical prediction (red dashed line) of $S_{x} = \qty{-172.7}{\decibelmilliwatt\per\hertz}$.}
    \label{fig:sensitivity}
\end{figure}

As established in \S~\ref{sec:calibration-system}, the calibrated data produced by the \ac{RFI} monitor are expressed as \acp{PSD}. As shown in \Cref{fig:sensitivity}, the observed noise floor of the instrument is $\approx\qty{-172}{\decibelmilliwatt\per\hertz}$. This sensitivity is several decibels better than high-end commercial real-time spectrum analyzers, such as the Tektronix RSA7100B or Rohde and Schwarz FSWX \citep{tektronixrsa7100B,rohdeschwarzfswx}, demonstrating the advantage of a signal chain specifically optimized for the \ac{DRAO} environment.

To validate this sensitivity, we calculated a refined estimate of the total system noise temperature, $T_{\rm sys}$. Combining the measured receiver temperature ($T_{\rm rec}=\qty{133.5}{\kelvin}$) from \S~\ref{sec:Trec-validation}, the predicted antenna temperature ($T_{\rm ant}=\qty{174}{\kelvin}$) from \S~\ref{sec:sys-temp-requirement}, and the contribution from the coaxial cable and adapters between the antenna and receiver ($T_{\rm coax}=\qty{84.8}{\kelvin}$), we find a system temperature of $T_{\rm sys} = T_{\rm ant} + T_{\rm coax} + T_{\rm rec} = \qty{392.3}{\kelvin}$. Applying the Johnson-Nyquist relation established in \S~\ref{sec:gain-stability-requirement}, this temperature corresponds to a predicted noise \ac{PSD} of \qty{-172.7}{\decibelmilliwatt\per\hertz}.  This prediction is in excellent agreement with the observed baseline $S_{x}$ shown in \Cref{fig:sensitivity}.

\section{Conclusions and Future Work} \label{sec:conclusion}

In this work, we have presented the design, implementation, and commissioning of the new wideband \ac{RFI} monitoring system at the \ac{DRAO}. The instrument now provides continuous, calibrated spectral coverage from \qtyrange{350}{1800}{\mega\hertz} with a system sensitivity of approximately \qty{-172}{\decibelmilliwatt\per\hertz} and stability sufficient for integrations exceeding 24 hours; a comprehensive summary of instrument specifications is provided in \Cref{tab:system-summary}. Through rigorous design and validation, the monitor achieves the precise temperature control, strict gain stability, low system noise temperature, and rapid calibration cadence required to support long-term statistical characterization of the radio environment at \ac{DRAO}.

The system significantly improves on the previous prototype in both performance and autonomy. The software-defined architecture, continuous operation, and standardized data interfaces allow it to serve as a shared infrastructure for spectrum management, instrument commissioning, and research into advanced \ac{RFI} detection algorithms. The ability to correlate \ac{RFI} events with site activities and environmental factors enables study of how local interference evolves over time and how it impacts scientific observations.

Planned upgrades will further expand the monitor’s capabilities. A dual-input \ac{IF} system will extend coverage to higher-frequency bands of interest (\qtyrange{2.75}{2.85}{\giga\hertz}, \qtyrange{8}{10}{\giga\hertz}, and the \qty{2.4}{\giga\hertz} unlicensed band), supporting instruments such as the \ac{SFM} and \ac{CGEM}. Additionally, an autonomous adaptive attenuation control loop will be developed to dynamically manage strong \ac{RFI} events while maintaining system linearity. On the software side, the real-time implementation of signal classification, modulation recognition, and anomaly detection will transform the monitor into a live research platform for machine-learning-based \ac{RFI} mitigation.

\begin{table}[htb!]
    \centering
    \caption{Summary of key system parameters for the commissioned \ac{DRAO} \ac{RFI} monitor.}
    \begin{tabular}{ll}
        \hline
        \textbf{Parameter} & \textbf{Value / Description} \\
        \hline
        Calibrated frequency coverage & \qtyrange{350}{1800}{\mega\hertz} (continuous) \\
        Instantaneous bandwidth & \qty{2}{\giga\hertz} \\
        Channel bandwidth & \qty{3.33}{\kilo\hertz} (standard; down to \qty{100}{\hertz} for long integrations) \\
        Integration time & \qty{0.75}{\second} (standard; down to \qty{50}{\milli\second} for transient characterization) \\
        System temperature & $\approx$\qty{390}{\kelvin} \\
        Receiver noise temperature & $\approx$\qty{133}{\kelvin} \\
        Gain stability & Continuous integration up to 24 hours without drift dominating \\
        On-sky fraction & $\geq$95\% \\
        Calibration cadence & \qty{270}{\second} \\
        Sensitivity & $\approx$\qty{-172}{\decibelmilliwatt\per\hertz} \\
        \hline
    \end{tabular}
    \label{tab:system-summary}
\end{table}

\begin{acknowledgments}
The authors gratefully acknowledge Elena Bergeron, Werner Dueck, Tayron Dueck, Ralph Webber, and Greg Nuspel for their help with the construction, mechanical design, and machining of the analog signal chain components. The authors thank the CHIME collaboration for supplying the data used to generate \Cref{fig:chime}.
\end{acknowledgments}

\begin{contribution}


NB conceptualized commissioning the prototype instrument, wrote software, contributed to the mechanical, electronic and \ac{RF} designs, performed formal analysis and validation, managed the project, constructed the instrument, wrote and submitted the manuscript.
CB contributed to the mechanical, electronic and \ac{RF} designs, was responsible for formal analysis and validation, project management, and editing the manuscript.
SH developed the initial prototype, did the \ac{FPGA} development, supervised NB, and edited the manuscript.
MI contributed to the mechanical design and edited the manuscript.
AO supervised employees, performed formal analysis and validation and edited the manuscript.
DL wrote software and edited the manuscript.
RM contributed to the mechanical, electronic and \ac{RF} designs, the instrument construction, and edited the manuscript.
BR contributed to the mechanical and electronic designs and the instrument construction.
TR performed formal analysis, and edited the manuscript.
PD supervised NB and edited the manuscript.

\end{contribution}

\facilities{DRAO}

\appendix

\section{Redesign Decisions} \label{app:redesign}

\subsection{Latching Switch and Matched Load}

\begin{figure}[htb!]
    \centering
    \includegraphics{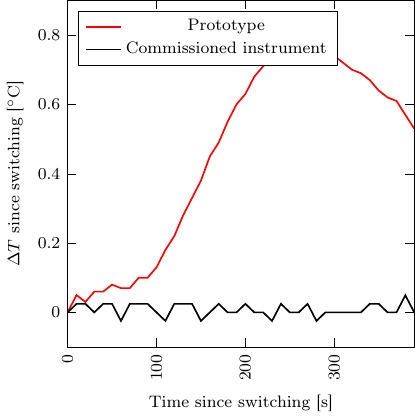}
    \caption{Matched-load temperature during switching before (red) and after (black) the redesign, showing that the matched load no longer heats up.}
    \label{fig:matched-load-during-switching}
\end{figure}

In the prototype, the antenna switch was not the latching type described in \S~\ref{sec:design-analog-receiver}. Instead, it dissipated power and generated heat when switched to the matched load. The matched load was a termination-style component that mounted directly onto the switch and had little thermal mass. As a result, the matched-load temperature would rise during the calibration, influencing the gain calibration results. \Cref{fig:matched-load-during-switching} shows the matched-load temperature during switching for both the old and new configurations, demonstrating that the matched-load temperature remains stable during switching because the switch is now latching and the new matched load has a much larger thermal mass tied into the copper backplane.

\subsection{Directional Coupler}

We also replaced the directional coupler, which injects the \qty{20}{\decibel} noise-diode calibration signal into the \ac{RF} chain. The original coupler had a low intrinsic insertion loss but occupied a large footprint and required SMA-to-N-type adapters at both ends to mate with the system's SMA cables. The replacement---a Tyco Electronics 2026-6012-20 coupler---features native SMA connectors and a smaller form factor, albeit with a higher intrinsic insertion loss. However, because the combined loss of the original coupler and its adapters exceeded that of the standalone Tyco unit, the new configuration yields a net performance gain. This reduction in overall signal path loss accounts for the slight improvement in $T_{\rm rec}$ shown in \Cref{fig:Trec}. Furthermore, the more compact signal chain fits into a smaller volume, making it easier to thermally stabilize.

\subsection{Mechanical Structure}

In the prototype, an approximately \qty{5}{\milli\meter}-thick aluminum sheet served as the thermal backplane, and an air-to-air \ac{TEA} interfaced with the external environment to control the signal-chain temperature. As \Cref{fig:matched-load-temp-tracking-ambient} shows, this system could not maintain a stable thermal environment. To resolve this stability issue, we redesigned the entire mechanical structure, incorporating a large copper block and the direct-to-air \ac{TEA} described in \S~\ref{sec:mechanical-structure}.

Furthermore, the loosely packed polystyrene foam insulation used in the original system was replaced with precision-cut Airex T92.60. This substantially reduces the air gap around the internal components.

\begin{figure}[htb!]
    \centering
    \includegraphics{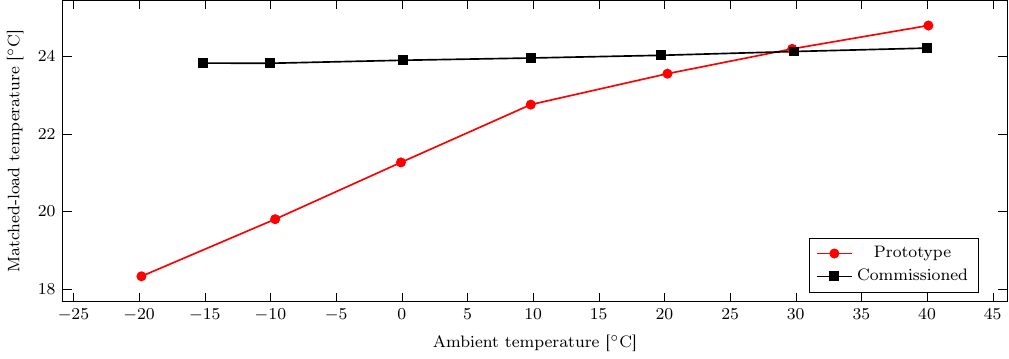}
    \caption{Matched-load temperature versus ambient air temperature. The mechanical redesign improves the thermal stability of the \ac{RF} signal chain by more than a factor of two.}
    \label{fig:ambient_vs_matched_load_temp}
\end{figure}

We tested the thermal stability of both the original and improved mechanical designs in an environmental chamber. \Cref{fig:ambient_vs_matched_load_temp} shows the matched-load temperature for both systems as the ambient temperature varied. In both tests, the \ac{TEA} had a temperature setpoint of \qty{24}{\celsius}.

\subsection{Receiver Noise Temperature}

\begin{figure}[htb!]
    \centering
    \includegraphics{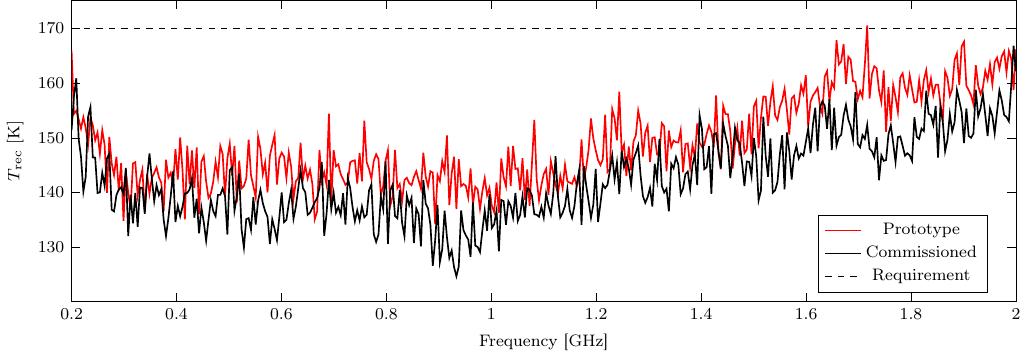}
    \caption{Receiver noise temperature for both the prototype (red) and the commissioned monitor (black) showing an $\approx\qty{10}{\kelvin}$ improvement.}
    \label{fig:Trec-comparison}
\end{figure}

The updated signal chain yields an $\approx\qty{10}{\kelvin}$ improvement in receiver noise temperature over the prototype, as shown in \Cref{fig:Trec-comparison}. We attribute this reduction in $T_{\rm rec}$ to the upgraded analog receiver components---specifically the new latching switch, the replacement directional coupler, and the elimination of the coaxial adapters.

\subsection{Temperature Signal Stability}

We monitor temperatures inside the signal-chain enclosure---specifically at the matched load, the noise diode, each \ac{LNA}, the copper surface, and the internal ambient air---using thermistors and the LabJack T4 digitizer. Originally, the unconditioned thermistor signals passed through D-subminiature \ac{RFI} filters on the \ac{RF}-shielded enclosure before reaching a resistor-divider conditioning circuit. However, testing revealed that these \ac{RFI} filters and a D-subminiature connector adapter coupled significant noise into the measurements. To resolve this, we relocated the conditioning circuit inside the signal-chain enclosure, ensuring that only the higher-level, conditioned voltages pass through the \ac{RFI} filters. This modification successfully eliminated the coupled noise.

\subsection{Temperature-Induced Gain Stability}

\begin{figure}[htb!]
    \centering
    \includegraphics{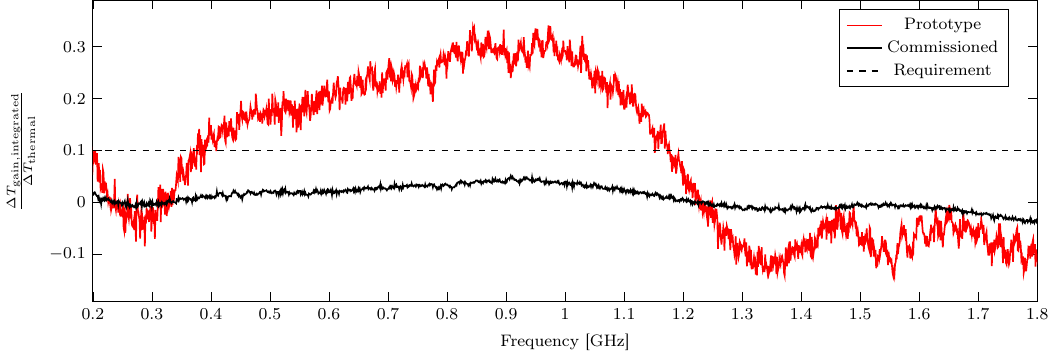}
    \caption{Gain stability comparison before (red) and after (black) changes were made to the system, showing that the requirement is now met for a \qty{1200}{\second} duration.}
    \label{fig:gain-stability-requirement-before-and-after}
\end{figure}

To determine if the prototype met the gain-stability requirement outlined in \S~\ref{sec:gain-stability}, we evaluated its performance to establish a baseline before implementing any hardware modifications. \Cref{fig:gain-stability-requirement-before-and-after} plots the integrated gain uncertainty defined by \Cref{eq:integrated-gain-uncertainty-requirement} to compare the system's performance before and after we redesigned the mechanical structure of the monitor.

\subsection{Calibration System}

The calibration system described in \S~\ref{sec:calibration-system} significantly reduced the calibration duration compared to the prototype. The original design relied on a networked LabJack T4 to control both the antenna-to-matched-load switch and the noise-diode switch. This reliance on network commands introduced timing uncertainties during switching events. To compensate, we had to extend the duration of each calibration stage (hot and cold) and discard the data acquired immediately before and after each transition. Consequently, the overall calibration process was inherently slow.

\section{Sky-Fraction Calculation} \label{app:sky-fraction}

In \S~\ref{sec:sys-temp-requirement}, we state that the monitor receives noise from an environment that is $\approx\qty{60}{\percent}$ ground and $\approx\qty{40}{\percent}$ sky. The local horizon has an average hill elevation angle of $\alpha = \qty{12.5}{\degree}$.

Let $R$ be the radius of a sphere, and let $\theta$ be the elevation angle from the equator. The radius of a horizontal slice at elevation $\theta$ is $R\cos\theta$, yielding a circumference of $2\pi R\cos\theta$. A differential strip at this elevation has an area $dA=(2\pi R\cos\theta)(Rd\theta) = 2\pi R^2\cos\theta d\theta$. Integrating this from the local horizon $\alpha$ to the zenith yields the exposed sky area:
\begin{equation} \label{eq:sky-area}
    A_\textrm{above}(\alpha) = \int^{\pi/2}_\alpha 2\pi R^2 \cos\theta d\theta = 2\pi R^2 (1-\sin\alpha)\,.
\end{equation}
Dividing the result of \cref{eq:sky-area} by the total surface area of a sphere, $A_\textrm{total}=4\pi R^2$, provides the fractional solid angle of the sky:
\begin{equation}
    \frac{A_\textrm{above}}{A_\textrm{total}} = \frac{1-\sin\alpha}{2}\,.
\end{equation}
Thus, for the measured average elevation angle $\alpha = \qty{12.5}{\degree}$, the fraction of the sky visible to the monitor is
\begin{equation}
    \frac{1-\sin(\qty{12.5}{\degree})}{2}=\qty{39.1}{\percent}\approx\qty{40}{\percent}\,.
\end{equation}

\section{Gain-Stability Algebra} \label{app:gain-stability-requirement-algebra}

We require the uncertainty attributable to the gain drift, $\Delta T_{\rm gain}$, to be no greater than 10\% of the total system noise uncertainty, $\Delta T_{\rm total}$. This condition is expressed as
\begin{equation}
    \Delta T_{\rm gain} \leq 0.1\Delta T_{\rm total}\,.
\end{equation}
Assuming that the thermal noise, $\Delta T_{\rm thermal}$, and the gain-drift noise are independent, they add in quadrature to produce the total uncertainty:
\begin{equation}
    \Delta T_{\rm total}^2 = \Delta T_{\rm thermal}^2 + \Delta T_{\rm gain}^2\,.
\end{equation}
Substituting the expression for total uncertainty into the initial requirement yields
\begin{equation}
    \Delta T_{\rm gain} \leq 0.1\sqrt{\Delta T_{\rm thermal}^2 + \Delta T_{\rm gain}^2}\,.
\end{equation}
Squaring both sides eliminates the radical and results in
\begin{equation}
    \Delta T_{\rm gain}^2 \leq 0.01 \left(\Delta T_{\rm thermal}^2 + \Delta T_{\rm gain}^2\right)\,.
\end{equation}
Distributing the fractional multiplier and isolating the $\Delta T_{\rm gain}$ terms on the left-hand side of the inequality, we find
\begin{equation}
    \Delta T_{\rm gain}^2 - 0.01 \Delta T_{\rm gain}^2 \leq 0.01 \Delta T_{\rm thermal}^2\,,
\end{equation}
which simplifies to
\begin{equation}
    0.99 \Delta T_{\rm gain}^2 \leq 0.01 \Delta T_{\rm thermal}^2\,.
\end{equation}
Dividing by 0.99 and taking the square root of both sides results in the upper bound:
\begin{equation}
    \Delta T_{\rm gain} \leq \frac{0.1}{\sqrt{0.99}} \Delta T_{\rm thermal}\,.
\end{equation}
Because $1/\sqrt{0.99} \approx 1.005$, setting the coefficient to 0.1 provides a conservative approximation of this upper limit accurate to within \qty{0.5}{\percent}:
\begin{equation}
    \Delta T_{\rm gain} \lesssim 0.1 \Delta T_{\rm thermal}\,.
\end{equation}
This inequality defines the strict error budget for the instrument: as long as temperature-induced gain fluctuations are suppressed to a full order of magnitude below the radiometric white noise, the system's total measurement uncertainty will remain fundamentally bounded by integration time rather than hardware instability.

\bibliography{bib}

 \end{document}